\documentclass[twocolumn,aps,prl,superscriptaddress]{revtex4-2}

\usepackage{amsmath}
\usepackage{graphicx}
\usepackage{MnSymbol}
\usepackage{parskip}

\begin{document}

\title{Sliding wear: role of plasticity}

\author{R. Xu}
\affiliation{State Key Laboratory of Solid Lubrication, Lanzhou Institute of Chemical Physics, Chinese Academy of Sciences, 730000 Lanzhou, China}
\affiliation{Peter Gr\"unberg Institute (PGI-1), Forschungszentrum J\"ulich, 52425, J\"ulich, Germany}
\affiliation{MultiscaleConsulting, Wolfshovener str. 2, 52428 J\"ulich, Germany}

\author{B.N.J. Persson}
\affiliation{State Key Laboratory of Solid Lubrication, Lanzhou Institute of Chemical Physics, Chinese Academy of Sciences, 730000 Lanzhou, China}
\affiliation{Peter Gr\"unberg Institute (PGI-1), Forschungszentrum J\"ulich, 52425, J\"ulich, Germany}
\affiliation{MultiscaleConsulting, Wolfshovener str. 2, 52428 J\"ulich, Germany}

\begin{abstract}
{\bf Abstract}: 
We present experimental wear data for polymethyl methacrylate (PMMA) sliding on tile, sandpaper, and polished steel surfaces, as well as for soda-lime, borosilicate, and quartz glass 
sliding on sandpaper. The results are compared with a recently developed theory \cite{ToBe} of sliding wear based on crack propagation (fatigue), 
originally formulated for elastic contact and here extended to include plasticity. 
The elastoplastic wear model predicts wear rates that agree reasonably well with the experimental results for PMMA
and soda-lime glass. However, deviations observed for quartz suggest that material-specific deformation mechanisms, particularly the differences between crystalline and amorphous structures, may need to be considered for accurate wear predictions across different materials. In addition, the model reveals a non-monotonic dependence of the wear
rate on the penetration hardness $\sigma_{\rm P}$. Thus, for plastically soft material, the wear rate increases with increasing $\sigma_{\rm P}$, while for hard materials, it decreases. This 
contrasts with Archard's wear law, where the wear rate decreases monotonically with increasing $\sigma_{\rm P}$.
\end{abstract}

\maketitle

\setcounter{page}{1}
\pagenumbering{arabic}




{\bf Corresponding author:} B.N.J. Persson, email: b.persson@fz-juelich.de
\vskip 0.3cm

{\bf 1 Introduction}


Wear is the progressive loss of material from a solid body due to its contact and relative movement against a 
surface \cite{wear1,wear2,Rabi1, Rabi2, Moli1, Moli2, Roland,Moli,W1,W2}. 
Wear particles can have an adverse influence on the health of living organisms, or result in 
the breakdown of mechanical devices.
There are several limiting wear processes, known as {\it fatigue wear}, 
{\it abrasive wear}, and {\it adhesive wear}. The actual wear process occurring depends on the surface roughness, the mechanical and chemical properties
of the material, and on external conditions such as sliding speed, load, temperature, and humidity.

Polymers are widely used in medical applications, but their wear can lead to serious complications.
For example, high-density polyethylene (HDPE) and ultra-high-molecular-weight polyethylene (UHMWPE) are used as bearing components in total joint replacements. 
During operation, wear particles generated from these materials can trigger inflammatory responses in surrounding tissue, potentially leading to osteolysis and implant loosening or failure \cite{inflame}. Similarly, polymethyl methacrylate (PMMA) is commonly used for prosthetic dental applications, such as artificial teeth, where wear may limit the functional lifetime of the component.

Understanding crack propagation is essential for analyzing wear. The fracture energy $G(v)$ is usually defined as the energy per unit (crack) surface area required to propagate a crack at a constant speed $v$. Here, we are concerned with crack growth under a time-dependent stress field, which results from interactions between asperities on the two contacting solids. In this context, one is interested in the crack growth function $\Delta x(\gamma)$, which describes the crack tip displacement $\Delta x$ during one oscillation of the driving stress, where the (maximum) elastic energy release rate equals $\gamma$. 

In the rubber community, $\gamma$ is commonly referred to as the tearing energy, defined as the energy per unit surface area required to create new crack surfaces. This concept was first introduced by Thomas \cite{Thomas0} in the context of rubber, where he observed a power-law relation $\Delta x \sim \gamma^n$ within a limited range of tearing energies. A similar relationship was later proposed by Paris for other materials \cite{Paris0}.

In the mechanical engineering community, instead of using $\gamma$, the displacement $\Delta x$ is often expressed as a function of the maximum stress intensity factor $K$ during the stress oscillation.

It is important to note that, in general, at least for viscoelastic solids such as rubber \cite{Persson0}, there is no simple relation between $G(v)$ and $\gamma$ (which depends on frequency), since $G(v)$ refers to the energy required to propagate a crack during steady crack growth, while $\gamma$ applies to oscillatory crack growth.

For polymers, the fracture or tearing energy $\gamma$ can range from $\sim 10^2$ to $\sim 10^5 \ {\rm J/m^2}$. This should be compared to the fracture energy of brittle crystalline solids, which is typically on the order of $\sim 1 \ {\rm J/m^2}$, even for materials with strong covalent bonds such as diamond. The large values of $\gamma$ observed in polymers result from additional energy dissipation mechanisms, including chain stretching, uncrosslinked chain pull-out, and mechanisms such as crazing, cavitation, and viscoelastic dissipation near the crack tip.

The fracture energy has been extensively studied for both constant crack tip velocity \cite{Gent} and oscillating strain conditions \cite{Rivlin, NatRub, Ghosh}, with both approaches yielding similar results. Under oscillatory strain, the crack tip displacement $\Delta x$ per strain cycle depends on $\gamma$. Below a critical threshold $\gamma_0$ (e.g., $\sim 50 \ {\rm J/m^2}$ for PMMA), no crack growth occurs, whereas $\Delta x$ diverges as $\gamma$ approaches the ultimate tear strength $\gamma_{\rm c}$ (e.g., $\sim 500 \ {\rm J/m^2}$ for PMMA). However, unless $\gamma$ is close to $\gamma_{\rm c}$, the crack tip displacement $\Delta x$ remains very small. As a result, multiple stress cycles may be required to remove a particle from a PMMA surface under conditions of {\it fatigue wear}.

Fracture energy is commonly characterized using macroscopic samples with characteristic dimensions on the order of $\sim 1 \ {\rm cm}$. However, this macroscopic scale may not accurately capture the fracture behavior relevant to polymer wear, which often involves material removal in the form of particles as small as $\sim 1 \ {\rm \mu m}$. At such small scales, the contributions of mechanisms like cavitation and crazing may be significantly reduced. Moreover, deformation in sliding contacts involves a broad distribution of loading frequencies, approximately given by $\omega \approx v / r_0$, where $v$ is the sliding speed and $r_0$ is the characteristic contact radius. This contrasts with the single-frequency loading conditions typically employed in standard tearing energy tests.

In one of the most widely used wear models, Archard \cite{A1,A3} proposed an empirical law stating that the wear volume is proportional to the normal 
load and sliding distance, and inversely proportional to the hardness of the material. This model, although simple, captures many of the trends observed in experiments. 
However, it does not provide a physical explanation for the underlying mechanisms of the wear process, and our study shows that the wear rate depends on the penetration hardness
in a different way than predicted by Archard's wear law.

Recently, in Ref. \cite{ToBe}, a theory was developed to describe sliding wear in elastic materials, where plastic deformation is negligible. The theory assumes that wear occurs in asperity contacts where the (temporarily) stored elastic energy is large enough to create the fracture surfaces involved in the removal of a wear particle. The model quantitatively predicts both the wear rate and the size of the wear particles, and shows good agreement with experimental results.

However, many materials such as polymers often exhibit significant plastic deformation in asperity contact regions. In this study, we extend the model to elastoplastic materials.

It is important to note that even if plastic flow occurs at some length scales, brittle fracture may still take place at larger length scales, so both mechanisms can operate simultaneously. To incorporate plastic deformation into the model, we make use of the following experimental observation: when two solids are pressed together with a normal force $F_{\rm N}$, some asperity contact regions may undergo plastic deformation. However, if the solids are then separated and brought back into contact at exactly the same position and with the same force $F_{\rm N}$, only elastic deformation occurs, and the resulting stress distribution is identical to that before separation. Therefore, the stress distribution predicted from the elastoplastic calculation is the appropriate one to use when estimating the elastic deformation energy.

In this study, we present wear rate measurements for PMMA sliding on tile, sandpaper, and polished steel surfaces under dry conditions, as well as for glass sliding on sandpaper, and compare the results with theoretical predictions.

A more detailed discussion of the historical development of wear models, with particular emphasis on soft materials, can be found in a recent review article \cite{history}.

\begin{figure}[!ht]
\includegraphics[width=0.48\textwidth]{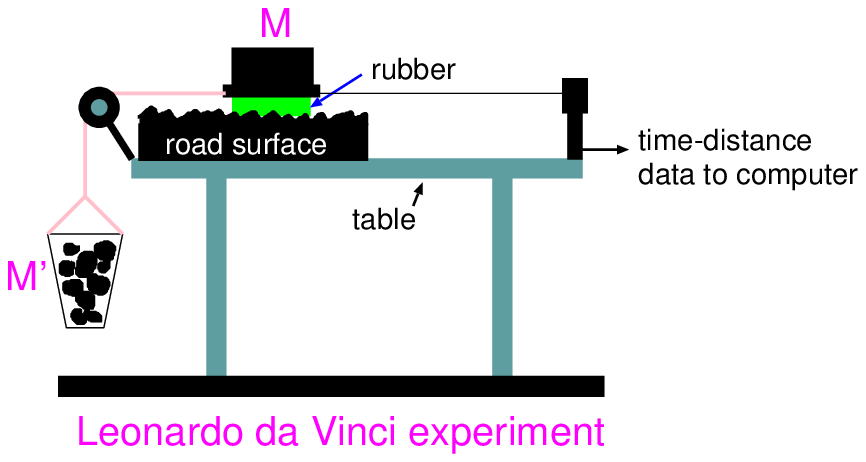}
\caption{\label{Leanardo.ps}
A simple friction slider (schematic) measures the sliding distance $x(t)$ via a displacement sensor.}
\end{figure}

\vskip 0.3cm
{\bf 2 Experimental methods}

The data used in the present study were obtained using the setup shown in Fig. \ref{Leanardo.ps}. The slider consists of a block (PMMA or glass) glued to a metal plate. The nominal contact area is $A_0 \approx 20 \ {\rm cm}^2$. The normal force $F_{\rm N}$ is determined by the total mass $M$ of lead blocks placed on top of the metal plate. Similarly, the driving force is controlled by the total mass $M'$ of lead blocks placed in the container. The PMMA block was tested on three different substrates: ceramic tile, sandpaper P100, and polished steel. Three types of glass: soda-lime (window glass), borosilicate, and quartz glass, were tested on sandpaper P100.

The sliding distance $x(t)$ as a function of time $t$ is measured using a displacement sensor. This simple friction slider setup can also be used to calculate the friction coefficient $\mu = M'/M$ as a function of sliding velocity and nominal contact pressure $p_0 = Mg/A_0$. Note that with this setup, the driving force is specified, allowing the study of the velocity dependence of friction only on the branch of the $\mu(v)$ curve where the friction coefficient increases with increasing speed. For the studied cases, the friction coefficient is very weakly velocity-dependent. Sometimes unstable sliding occurs, resulting in a sliding speed that fluctuates over time. The average sliding speed in our studies was $\sim 3 \ {\rm mm/s}$.

To study the velocity and pressure dependence of the wear rate, we slid the metal plate-block system on the tested surfaces for different distances: $21.5 \ {\rm cm}$ on tile, $18 \ {\rm cm}$ on sandpaper, and $10 \ {\rm cm}$ on polished steel. The wear rate was determined from the mass change, defined as the difference in the mass of the plate-block system before and after sliding, using a high-precision balance (Mettler Toledo analytical balance, model MS104TS/00) with a sensitivity of $0.1 \ {\rm mg}$. After each sliding sequence, the surface was cleaned using a brush or a single-use nonwoven fabric.

The surface roughness of all surfaces used in this study was measured using a Mitutoyo Portable Surface Roughness Measurement Surftest SJ-410. The instrument is equipped with a diamond tip with a radius of curvature of $R = 1 \ {\rm \mu m}$ and operates with a tip-substrate repulsive force of $F_N = 0.75 \ {\rm mN}$. Measurements were taken with a step length (pixel) of $0.5 \ {\rm \mu m}$, a scan length of $L = 25 \ {\rm mm}$, and a tip speed of $v = 50 \ {\rm \mu m/s}$.

\begin{figure}
\includegraphics[width=0.47\textwidth,angle=0.0]{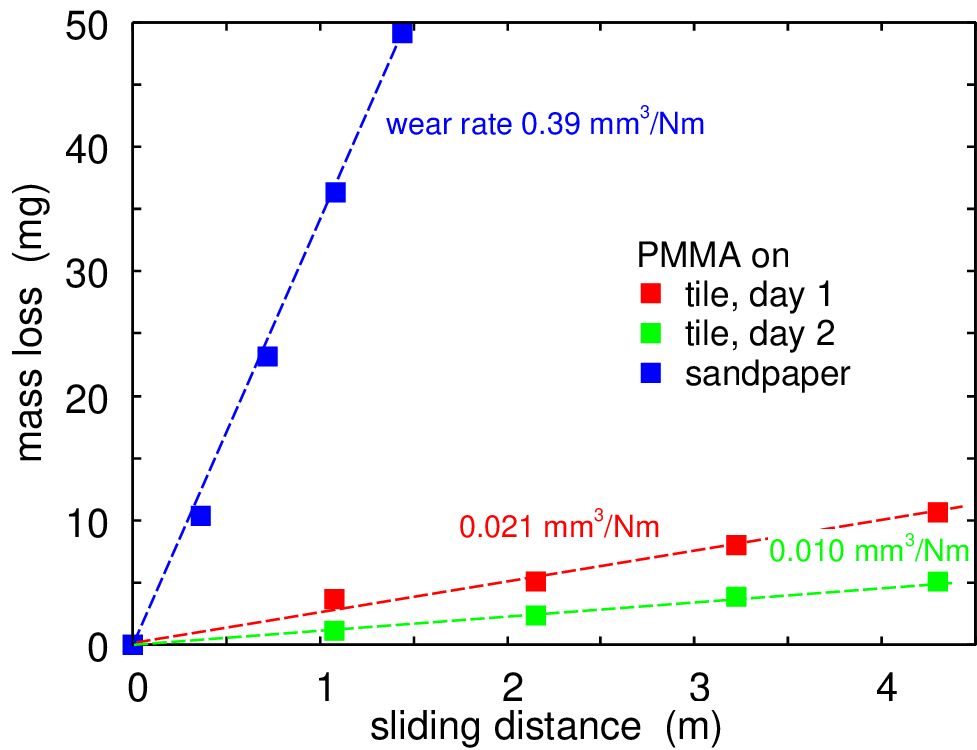}
\caption{\label{1distance.massloss.PMMA.tile.and.sandpaper.eps}
The mass loss of the PMMA blocks as a function of the sliding distance on the tile and sandpaper P100 surface. 
Experiments on the tile surface performed on two consecutive days gave different wear rates.
}
\end{figure}

\begin{figure}
\includegraphics[width=0.47\textwidth,angle=0.0]{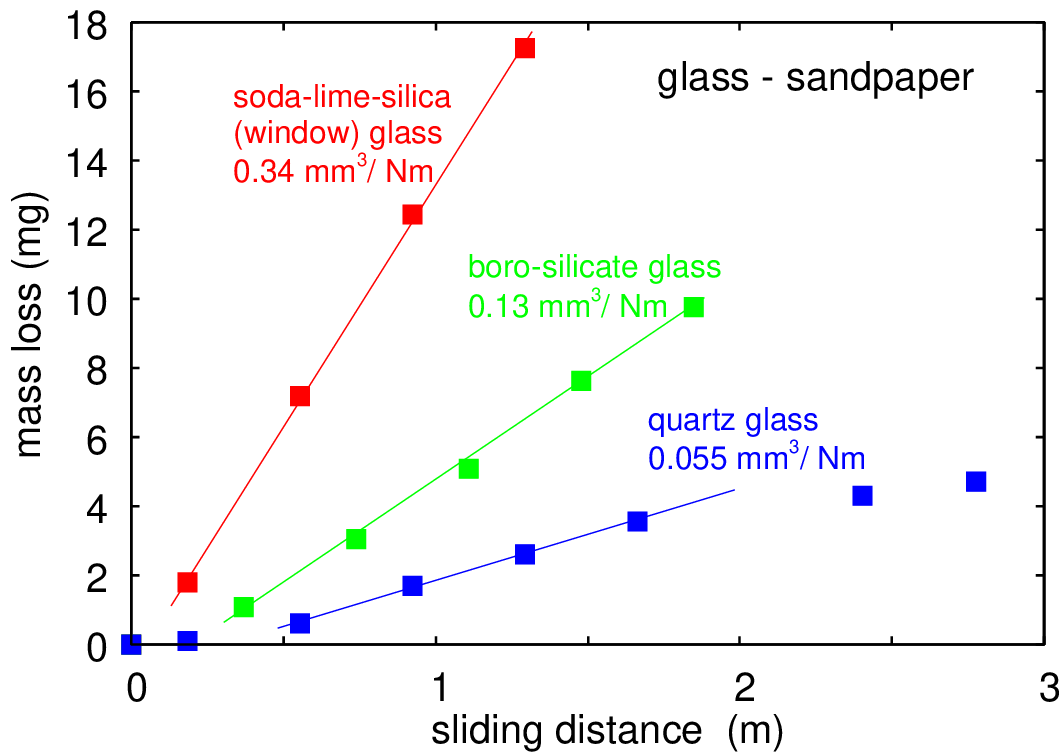}
\caption{\label{1distance.2massloss.allGLASS.eps}
The mass loss of blocks made of soda-lime glass (red), borosilicate glass (green), and quartz glass as a function of the sliding distance. The substrate used is sandpaper P100. The wear rates, $\Delta V/F_{\rm N}L$, are indicated in the figure. 
}
\end{figure}

\begin{figure}
\includegraphics[width=0.47\textwidth,angle=0.0]{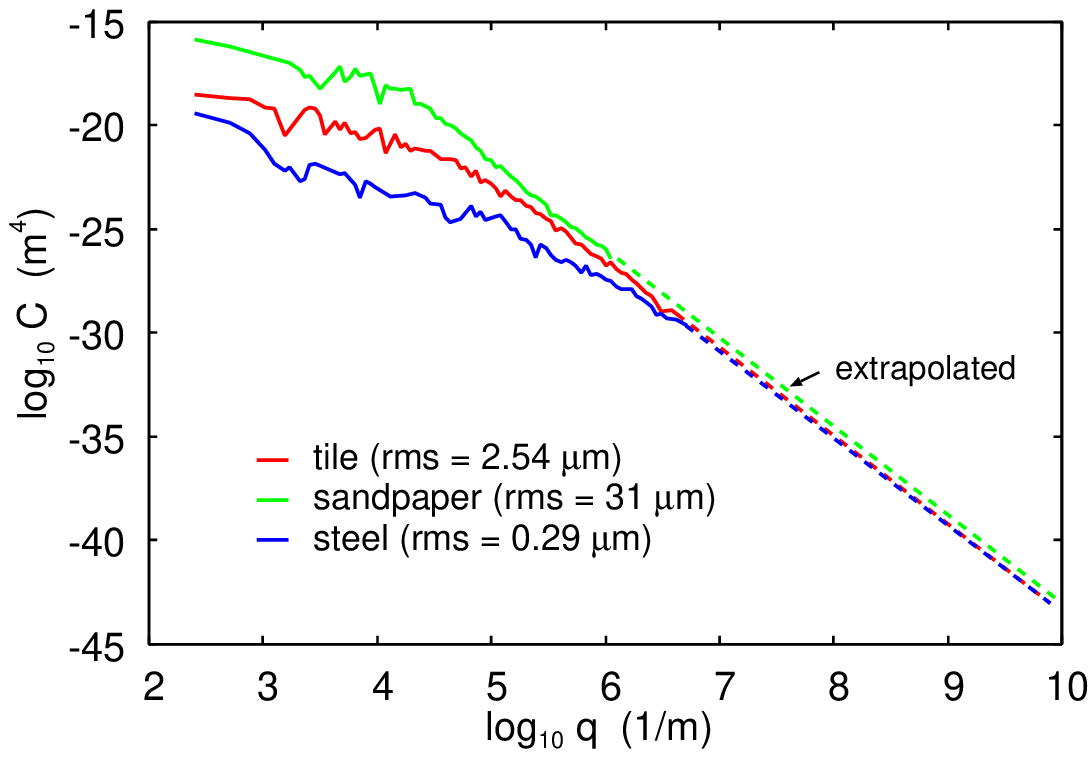}
\caption{\label{1logq.2logC.tile.steel.P100.eps}
The surface roughness power spectra of the tile, steel and sandpaper surfaces used in the wear studies.
The dotted regions are extrapolated with a slope corresponding to the Hurst exponent $H\approx 1$.
}
\end{figure}

\vskip 0.3cm
{\bf 3 Experimental results}

We have measured the wear rate and the friction force of PMMA and three types of glass 
blocks sliding on the tested surfaces. In all experiments, the normal load was $F_{\rm N} = 104 \ {\rm N}$, and the nominal contact area was $A_0 = 20 \ {\rm cm^2}$, resulting in a nominal contact pressure of $p_0 = 0.052 \ {\rm MPa}$. 
Since PMMA, glass, ceramic tile, sandpaper, and polished steel are all relatively stiff materials, the blocks do not make uniform contact with the substrate surfaces at the macroscopic level. 
This non-uniformity is evident from the wear track patterns observed on the surface of blocks after a sliding act. 
As a result, the nominal contact pressure is not uniform but varies on the length scale of the block dimensions.

The green and red lines and symbols in Fig. \ref{1distance.massloss.PMMA.tile.and.sandpaper.eps} show the mass loss of the 
PMMA blocks as a function of the sliding distance for the tile surface.
Experiments performed on two consecutive days gave different results: on the first day, the wear rate was $\Delta V/F_{\rm N}L = 0.021 \ {\rm mm^3/Nm}$, while on the second day, it was $0.010 \ {\rm mm^3/Nm}$. Here, we calculated the wear volume from the mass loss, assuming a PMMA mass density of $\rho = 1180 \ {\rm kg/m^3}$.
In both measurements, the wear rate was proportional to the sliding distance, suggesting that the difference in wear rate between the two days must be due to 
changes in experimental conditions (e.g., humidity, which was not measured) or some aging process that altered the properties of the worn PMMA or tile surface. 
The PMMA block was cut from a PMMA sheet, and during the cutting process, the surface was flooded with a cutting fluid (a water–oil emulsion). It is known that PMMA absorbs water, which acts as a plasticizer and reduces the penetration hardness. This, in turn, may increase the wear rate due to an increase in the real contact area. Since the experiments were performed shortly after the PMMA block was prepared, the penetration hardness may have been lower on the first day compared to the second day. We did not observe any PMMA particles adhering to the tile surface; however, we cannot exclude the possibility that the asperities on the tile surface were covered by a nanometer-thin film of PMMA.

The blue line and symbols in Fig. \ref{1distance.massloss.PMMA.tile.and.sandpaper.eps} show the mass loss of the PMMA block on the sandpaper 
P100 surface, where a significantly higher wear rate was observed, with $\Delta V/F_{\rm N}L = 0.39 \ {\rm mm^3/Nm}$.
In contrast, for the polished steel surface, no mass change was detected after a sliding distance of $4 \ {\rm m}$. 
However, given that the resolution of the measuring instrument was $0.1 \ {\rm mg}$, it is possible that a smaller amount of 
wear may have occurred but was below the detection limit.

The wear produced a white powder consisting of PMMA particles, which was easily brushed away after each sliding act. The PMMA wear particles were transparent, and the optical method available to us did not provide sufficient contrast. As a result, we were unable to analyze the size of the PMMA wear particles using the same optical microscope that was used for rubber wear particles in Ref. \cite{ToBe}.

The wear rate for glass blocks sliding on sandpaper P100 surfaces is shown in Fig. \ref{1distance.2massloss.allGLASS.eps}. The figure presents the mass loss of blocks made of soda-lime (red), borosilicate (green), and quartz glass as a function of the sliding distance. The corresponding wear rates, given by $\Delta V/F_{\rm N}L = 0.34$, $0.13$, and $0.055 \ {\rm mm^3/Nm}$ for the three glass types, respectively, were calculated using assumed mass densities of $\rho = 2440$, $2230$, and $2200 \ {\rm kg/m^3}$ for soda-lime, borosilicate, and quartz glass, respectively. As in the case of PMMA, we were unable to analyze the wear particles using optical methods due to their transparency and equipment limitations.

Fig. \ref{1logq.2logC.tile.steel.P100.eps} shows the surface roughness power spectra of the tile, sandpaper, and steel surfaces used in the wear studies. The dotted regions represent extrapolated portions with a slope corresponding to a Hurst exponent of $H \approx 1$. Both the tile and steel surfaces are harder and elastically stiffer than PMMA, and are therefore treated as rigid with no deformation of their surface roughness profiles. The sandpaper surface consists of very hard corundum (aluminum oxide) particles, which can also be considered rigid when in contact with PMMA and even with glass surfaces. (The penetration hardness of corundum is approximately $30 \ {\rm GPa}$, compared to about $15 \ {\rm GPa}$ for quartz.)

\begin{figure}
\includegraphics[width=0.3\textwidth,angle=0.0]{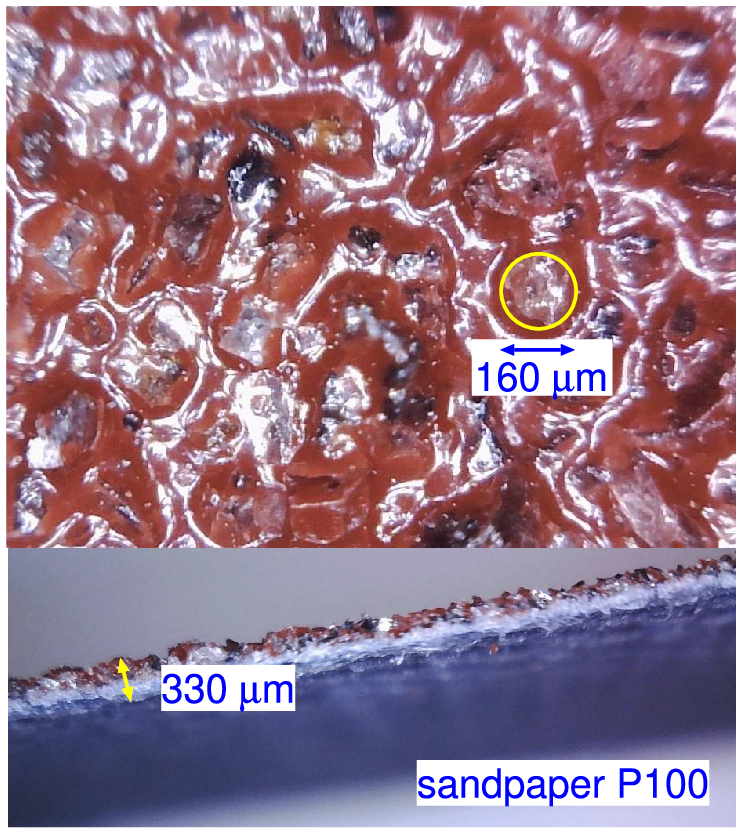}
\caption{\label{SandPaperP100.ps}
Top: Picture of sandpaper P100. The corundum particles have an average diameter of $\approx 160 \ {\rm \mu m}$.
Bottom: Cross-section of the sandpaper. The sandpaper P100 has a thickness of $\approx 330 \ {\rm \mu m}$.
}
\end{figure}

However, the corundum particles are embedded in a polymer fiber mat that contains an acrylic resin (see Fig. \ref{SandPaperP100.ps}), which is elastically and plastically much softer than glass and likely similar in properties to PMMA. When sandpaper is pressed against silica glass surfaces, surface roughness components with wavelengths longer than the size of the corundum particles (diameter $D \approx 160 \ {\rm \mu m}$ for P100 sandpaper; see Fig. \ref{SandPaperP100.ps}) are easily flattened. These long-wavelength components should not be included in wear rate calculations for silica glass surfaces if the substrate is treated as rigid. For this reason, we exclude the roughness components with wavenumbers $q < 2\pi /D \approx 4 \times 10^4 \ {\rm m^{-1}}$ from the power spectrum of the sandpaper surface (green line in Fig. \ref{1logq.2logC.tile.steel.P100.eps}) when calculating wear rates for glass surfaces. For PMMA on sandpaper, it is less clear whether this same power spectrum correction is necessary. However, for consistency, we apply the same long-wavelength cut-off for PMMA in the present study.

In this study we neglected the surface roughness of the PMMA and glass surfaces because, in all cases, it was much smaller than that of the hard countersurfaces.

\begin{figure}
\includegraphics[width=0.25\textwidth,angle=0.0]{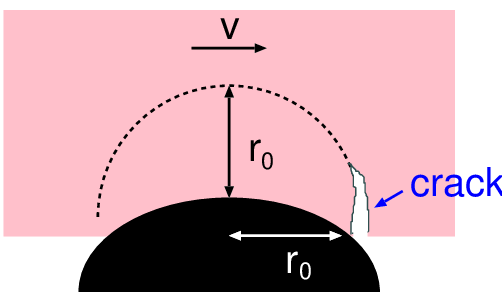}
\caption{\label{WearParticlePMMA.eps}
A PMMA block sliding in contact with a hard countersurface. 
The sliding speed $v$ and the radius of the contact 
region $r_0$ are indicated. 
The deformation field extends into the polymer a similar distance as it extends laterally.
}
\end{figure}

\vskip 0.3cm
{\bf 4 Theory of sliding wear}

Sliding wear depends on the size of the contact regions and on the stress acting within these regions \cite{Mu1}. The theoretical model used in this study follows the approach introduced in Ref. \cite{ToBe} for rubber wear. Here, we present an alternative derivation of the main result, which is also extended to include plasticity effects relevant to PMMA/glass sliding contacts.

Cracks at the surface of a solid
can be induced by both the normal and the tangential stress acting on the surface,
but particle removal is caused mainly by the tangential stress.
Let $\tau = \tau (\zeta_r)$ be the effective 
shear stress acting in an asperity contact region with a radius $r_0$.
The magnification $\zeta_r$ is determined by the radius of the contact region,
$q_r = \pi /r_0$,  $\zeta_r = q_r /q_0$.
The elastic energy stored in the deformed asperity contact is (see Fig. \ref{WearParticlePMMA.eps})
$$U_{\rm el} \approx {\tau^2  \over  E^*} r_0^3,$$
where the effective modulus $E^* = E/(1-\nu^2)$ (we assume that the substrate is rigid).
More accurately, assume that the shear stress acts uniformly within a circular region
with a radius $r_0$. The center of the circular region will displace a distance $u$ given by
$k u = F$, where $F=\tau \pi r_0^2$ is the force and $k \approx (\pi/2) E^* r_0$ the spring constant.
This gives the elastic energy
$$U_{\rm el} = {1\over 2} k u^2 = {F^2 \over 2 k} = 
{(\pi r_0^2 \tau)^2\over \pi E^* r_0} = \pi {\tau^2 \over E^*} r_0^3.\eqno(1)$$
In order for the shear stress to
remove a particle of linear size $r_0$, the stored elastic energy must be larger than the
fracture (crack) energy, which is of the order
$$U_{\rm cr} \approx \gamma 2 \pi r_0^2, \eqno(2)$$
where $\gamma$ is the energy 
per unit surface area to break the bonds at the crack tip. If $U_{\rm el}>U_{\rm cr}$,
the elastic energy is large enough to propagate a crack and
remove a particle\cite{Rabi1,Rabi2,Moli1}. Thus, for a particle to be 
removed, we must have $\tau > \tau_{\rm c}$, where
$$\tau_{\rm c} = \beta \left ({2 E^* \gamma \over r_0 }\right )^{1/2}, \eqno(3)$$
where $\beta$ is a number of order unity, which takes into account that the wear particles, in general,
are not hemispherical as assumed above.

In what follows, we will treat the polymer surface as 
smooth and assume only roughness on the counter surface.
We will denote a substrate asperity, where the shear stress is high enough to remove a particle
of size $r_0$,  as a {\it wear-asperity}, and the corresponding contact region as the {\it wear-contact region}.
 
If we assume that during sliding, the effective 
shear stress $\tau$ is proportional to the normal stress $\sigma$, $\tau = \mu \sigma$,
we find that particles will be removed only if the contact stress $\sigma > \sigma_{\rm c} (\zeta)$, where
$$\sigma_{\rm c} = {\beta \over \mu} \left ({2 E^* \gamma \over r_0 }\right )^{1/2}. \eqno(4)$$

For randomly rough surfaces, for elastic contact the probability distribution of contact stress equals:
$$P(\sigma,\zeta) = {1\over (4\pi G)^{1/2}} \left (e^{-(\sigma-\sigma_0)^2/4G}- e^{-(\sigma+\sigma_0)^2/4G} \right ), \eqno(5)$$
where $\sigma_0$ is the nominal (applied) pressure and where
$$G ={\pi \over 4} (E^*)^2 \int_{q_0}^{\zeta q_0} dq \ q^3 C(q), \eqno(6)$$
where $C(q)$ is the surface roughness power spectrum.

\begin{figure}
\includegraphics[width=0.3\textwidth,angle=0.0]{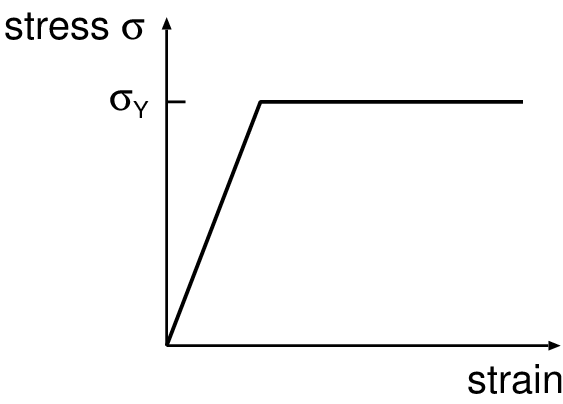}
\caption{\label{Plasticity.eps}
The relation between the stress and the strain in elongation for the simplest elastoplastic model 
assumes that the maximum stress equals the yield stress, $\sigma_{\rm Y}$. The penetration hardness 
is typically $\sigma_{\rm P} \approx 3 \sigma_{\rm Y}$, where $\sigma_{\rm P}$ represents the ratio 
between the indentation force and the indentation cross-sectional area.
}
\end{figure}

When the stress in the asperity contact region becomes high enough, plastic flow occurs. In the simplest model, it is assumed that a material deforms as a linear elastic solid until the stress reaches a critical level, the so-called plastic yield stress, where it flows without strain hardening (see Fig. \ref{Plasticity.eps}). The yield stress in elongation is denoted by $\sigma_{\rm Y}$. In indentation experiments, where a sharp tip or a sphere is pushed against a flat solid surface, the penetration hardness $\sigma_{\rm P}$ is defined as the ratio between the normal force and the projected (on the surface plane) area of the plastically deformed indentation. Typically, $\sigma_{\rm P} \approx 3 \sigma_{\rm Y}$.
We note that the yield stress of materials often depends on the length scale (or magnification) which in principle can be included in the
formalism we use \cite{Preview,Brodsky}.

The influence of plastic flow on the contact mechanics can be taken into account in the Persson contact mechanics approach by replacing the boundary condition $P(\infty,\zeta) = 0$ with the condition that there is no stress at the interface above the penetration hardness, i.e., $P(\sigma, \zeta) = 0$ for $\sigma > \sigma_{\rm P}$. Thus, the maximum stress at the interface is equal to the penetration hardness $\sigma_{\rm P}$. This approach is based on the simplest elastoplastic description, where only elastic deformation occurs for $\sigma < \sigma_{\rm P}$, while for $\sigma = \sigma_{\rm P}$, the material flows without work-hardening so that the maximal stress equals $\sigma_{\rm P}$ (see Fig. \ref{Plasticity.eps}). The pressure probability distribution for this case is given by\cite{JCPP}:
$$P(\sigma,\zeta) = {2\over \sigma_{\rm P}} \sum_{n=1}^\infty \, {\rm sin} (s_n\sigma_0) \, {\rm sin} (s_n\sigma) \, e^{-s_n^2 G(\zeta)}$$
$$+P_{\rm pl} (\zeta) \delta (\sigma-\sigma_{\rm P})\eqno(7)$$
where $s_n = n \pi/\sigma_{\rm P}$ and
$$P_{\rm pl} = {\sigma_0\over \sigma_{\rm P}}+{2\over \pi} \sum_{n=1}^\infty \, {(-1)^n \over n} 
{\rm sin} (s_n\sigma_0) \, e^{-s_n^2 G(\zeta)}\eqno(8)$$

As $\sigma_{\rm P} \rightarrow \infty$, (7) reduces to (5). The $P(\sigma,\zeta)$ is also the pressure distribution resulting from {\it elastic} deformations if the two surfaces are separated and brought into contact again at the same position. Hence, it is the pressure distribution that should be used to obtain the elastic energy, which enters into the theory of the wear rate.

\begin{figure}
\includegraphics[width=0.45\textwidth,angle=0.0]{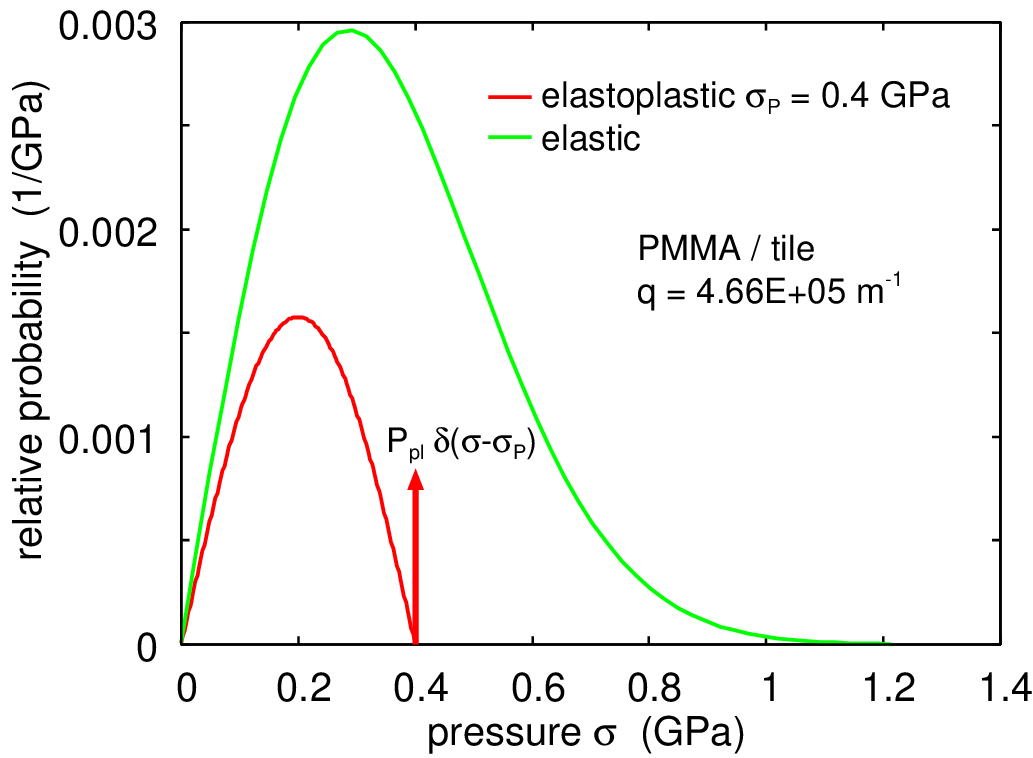}
\caption{\label{1pressure.2Probability.elast.and.plast.q=4.66E5.eps}
The stress probability distribution $P(\sigma,\zeta)$ as a function of the stress $\sigma$ for PMMA in contact with
the tile surface, with the power spectrum given by the red curve in Fig. \ref{1logq.2logC.tile.steel.P100.eps}.
The magnification $\zeta = q/q_0$, with $q = 4.66 \times 10^5 \ {\rm m}^{-1}$ and $q_0 = 251 \ {\rm m}^{-1}$. The tile surface is considered
as rigid and the PMMA as elastic (green curve) with $E = 3 \ {\rm GPa}$ and $\nu = 0.3$, or elastoplastic (red curve) with 
$\sigma_{\rm P} = 0.4 \ {\rm GPa}$.
}
\end{figure}

In Fig. \ref{1pressure.2Probability.elast.and.plast.q=4.66E5.eps}, we show $P(\sigma,\zeta)$ as a function of the stress $\sigma$ for $\zeta = 1857$ for PMMA in contact with the tile surface, with the power spectrum given by the red curve in Fig. \ref{1logq.2logC.tile.steel.P100.eps}. The tile surface is considered as rigid and the PMMA elastic (green curve) or elastoplastic (red curve) with the penetration hardness $\sigma_{\rm P} = 0.4 \ {\rm GPa}$. 

When the interface is studied at the magnification $\zeta$, 
the area $A=A_{\rm wear}(\zeta)$, where the shear stress is high enough to remove particles, is given by
$${A_{\rm wear} (\zeta) \over A_0} = \int_{\sigma_{\rm c} (\zeta)}^\infty d\sigma \ P(\sigma,\zeta). \eqno(9)$$

When we study the interface at the magnification $\zeta$, the smallest wear particles observed have the
size $r_0 \approx \pi /q_r$, with $q_r=\zeta q_0$. 
We may say that at the magnification $\zeta$, there is a pixel size of $r_0 =\pi /\zeta q_0$, 
and the smallest removed particle, which
can be observed at this magnification, is determined by the pixel size.
As previously stated, such a particle can be removed from the polymer surface if $U_{\rm el} > U_{\rm cr}$,
where $U_{\rm el}$ is the stored elastic energy ($\sim r_0^3$) in a volume element of linear size $r_0$, and $U_{\rm cr}$ is the energy
needed to break the bonds and detach the particle. $U_{\rm cr}\approx \gamma 2\pi r_0^2$, where $\gamma$ is the energy per unit surface area to propagate the crack. The crack energy $\gamma$ 
depends on the speed of bond-breaking and will take a range of values,
$\gamma_0 < \gamma < \gamma_{\rm c}$. The faster the crack
propagates, the larger $\gamma$ becomes. The smallest stored elastic energy $U_{\rm el}= U_{\rm el 0}$, which can remove a particle, is
given by $U_{\rm el 0} \approx \gamma_0 2\pi r_0^2$, but for this case, the crack moves extremely slowly, and the incremental displacement $\Delta x$
during the interaction between the (moving) {\it wear-asperity} and the crack is very small, requiring many $\sim r_0/\Delta x$ contacts to remove the particle. If the interaction with a {\it wear-asperity} results in $U_{\rm el}>> U_{\rm el 0}$, 
the crack will move much faster ($\Delta x$ is much bigger), and far fewer contacts are needed to remove a particle.
During sliding, the crack will be in contact with many {\it wear-asperities} of different sizes, 
so it will experience a wide range of crack-tip movements $\Delta x$ before the particle is finally removed.

The probability that the stress at an arbitrary point on the polymer surface is between $\sigma$ and $\sigma+d\sigma$, when the interface
is studied at the magnification $\zeta$, is given 
by $P(\sigma,\zeta) d\sigma$. If $\sigma > \sigma_{\rm c}$, the local stress
results in $U_{\rm el}> U_{\rm el 0}$, so in principle, a particle could be removed. But during the interaction time, the crack moves
only the distance $\Delta x (\gamma)$, where we assume the relevant $\gamma$ is given by $U_{\rm el}=\gamma r_0^2$.
Hence, $N(\gamma) = r_0/\Delta x$ contacts are needed to remove the particles. Thus, after the run-in, the probability that a particle will be removed
from the regions where the stress is in the range $\sigma$ to $\sigma+d\sigma$
will be $P(\sigma,\zeta) d\sigma /N(\gamma)$. The total probability will be
$$P^*=\int_{\sigma_{\rm c}}^\infty d\sigma {P(\sigma, \zeta) \over 1+ r_0(\zeta )/\Delta x(\sigma, \zeta)}\eqno(10)$$
where we have added 1 in the denominator in order for the limit 
$\Delta x/r_0 \rightarrow \infty$ to be correct.
In (10), the cut-off stress $\sigma_0$ is determined by $U_{\rm el}= U_{\rm el 0}$.
There are $N^* = A_0/\pi r_0^2$ pixels on the surface, so sliding the distance $L=2r_0$ 
will result in removing $N^* P^*$ particles,
corresponding to the volume $V = (2 \pi r_0^3/3) N^* P^*$. Thus, we get $V/L = (\pi r_0^2/3) N^* P^*$ or
$V/LA_0 = P^*/3$. Using (10), this gives
$${V\over LA_0}={1\over 3} \int_{\sigma_{\rm c}}^\infty d\sigma {P(\sigma, \zeta) \over 1+ r_0(\zeta )/\Delta x(\sigma, \zeta)}\eqno(11)$$
which is the same as (17) in Ref. \cite{ToBe},
except that the factor of 1/2 in (17) in Ref. \cite{ToBe} is replaced by 1/3 in (10) due to a slightly
different description of the particle removal process.
Eq. (11) shows that the wear volume per unit sliding length is proportional to the nominal surface area, as expected
when the nominal contact pressure is constant.

The number of contacts needed to remove a particle $N_{\rm cont} \approx r_0/\Delta x$ 
depends on the crack energy $\gamma$, but it could be a large number ($10^2$ or more) 
if the macroscopic relation between the tear-energy 
$\gamma$ and $\Delta x$ would also hold at the length scale of the
wear particles. 

The theory above estimates the wear volume by considering particles of a specific size characterized by radius $r_0$. At the magnification level $\zeta = q_r/q_0 = \pi / q_0 r_0$, these correspond to the smallest observable wear particles. To obtain the total wear volume, contributions from all relevant length scales must be accumulated as the magnification increases. To avoid double-counting of similarly sized particles, the magnification is incremented in steps of approximately a factor of 2, expressed as $\zeta = 2^n = \zeta_n$, where $n = 0, 1, \dots, n_1$ and $2^{n_1} q_0 = q_1$. Each range between $\zeta = 2^n$ and $2^{n+1}$ is referred to as a two-interval.
Using
$$\sum_{n=0}^{n_1} f_n \approx \int_0^{n_1} dn \ f_n = 
{1\over {\rm ln}2} \int_1^{\zeta_1} d\zeta \ {1\over \zeta} f(\zeta),$$ 
we can write the total wear volume when $\Delta x$ is constant as
$${V\over A_0 L} \approx  {1\over 3}
\sum_{n=0}^{n_1} {1\over 1+r_0(\zeta_n)/\Delta x} {A_{\rm wear} (\zeta_n) \over A_0}$$
$$\approx {1\over 3 {\rm ln}2 }  \int_1^{\zeta_1} d\zeta \ {1\over \zeta} {1\over 1+r_0(\zeta)/\Delta x} {A_{\rm wear} 
(\zeta) \over A_0}. \eqno(12)$$ 
Using $\zeta r_0 = \pi/q_0$, this gives
$${V\over A_0 L} \approx  
{1\over 3  {\rm ln}2 }  \int_1^{\zeta_1} d\zeta \ {1\over \zeta+\pi/q_0\Delta x} {A_{\rm wear} (\zeta) \over A_0}$$
$$={1\over 3 {\rm ln}2 } \int_{q_0}^{q_1} dq  \ {1\over q+\pi/\Delta x} {A_{\rm wear} (q) \over A_0}. \eqno(13)$$
When $\Delta x$ depends on $\gamma$, we get
$${V\over A_0 L} ={1\over 3 {\rm ln}2 } \int_{q_0}^{q_1} dq  
\int_{\sigma_{\rm c} (\zeta)}^\infty d\sigma \ 
{P(\sigma,\zeta) \over q+\pi/\Delta x(\sigma,\zeta)}, \eqno(14)$$
where $\zeta = q/q_0$. It is convenient to write $q=q_0 e^\xi$, so that $dq=q d\xi$, and
$${V\over A_0 L} ={1\over 3 {\rm ln}2 } \int_{0}^{\xi_1} d\xi  
\int_{\sigma_{\rm c} (\zeta)}^\infty d\sigma \ 
{P(\sigma,\zeta) \over 1+r_0(\zeta)/\Delta x(\sigma,\zeta)} .\eqno(15)$$
where $\xi_1 = {\rm ln}(q_1/q_0)$.

If we write 
$$Q(\sigma,\zeta) = {P(\sigma,\zeta) \over 1+r_0(\zeta)/\Delta x(\sigma,\zeta)}$$
we can define the average number of contacts needed to remove a particle of size $r_0=\pi/\zeta q_0$
as
$$\langle N_{\rm cont} \rangle = {\int_{\sigma_{\rm c} (\zeta)}^\infty d\sigma \ N_{\rm cont}(\sigma,\zeta) Q(\sigma,\zeta) \over
\int_{\sigma_{\rm c} (\zeta)}^\infty d\sigma \ Q(\sigma,\zeta) }\eqno(16)$$
where
$$N_{\rm cont}(\sigma,\zeta) = 1+r_0(\zeta)/\Delta x(\sigma,\zeta) .$$

The distribution of particles of different sizes is given by (17) [or (18)]. Thus, the number of
particles with radius $r_0$ between $(\pi / q_0) 2^{-n-1/2}$ and $(\pi / q_0) 2^{-n+1/2}$ is
$${N_n \over A_0 L} \approx {1 \over 3 \pi r_0^3 (\zeta_n) [1+r_0 (\zeta_n)/\Delta x]} {A_{\rm wear} (\zeta_n) \over A_0}\eqno(17)$$
or, when $\Delta x$ depends on $\gamma$,
$${N_n \over A_0 L} \approx {1 \over 3 \pi r_0^3 (\zeta_n) } 
\int_{\sigma_{\rm c} (\zeta_n)}^\infty d\sigma \ 
{P(\sigma,\zeta_n) \over 1+r_0(\zeta_n)/\Delta x(\sigma,\zeta_n)}. \eqno(18)$$

The theory presented above assumes that all length scales contribute independently to the wear rate.
This cannot be strictly true since a long crack, which would result in a large wear
particle, will change the stress field in its vicinity out to a distance of the order of the length of the crack.
This effect, known as crack shielding, 
reduces the ability for smaller cracks 
to grow in the neighborhood of longer cracks. However, crack tip shielding is 
much weaker for sliding contacts as compared to polymer strips elongated by uniform 
far-field stress.

Note that if $r_0/\Delta x$ is large, a long run-in distance would be needed before the wear reaches a steady state.
This is particularly true if the nominal contact pressure is small, where the distance between the wear asperity
contact regions may be large. However, since the contact regions within the macroasperity contacts are densely distributed
and independent of the nominal contact pressure, there may, in some cases, be enough wear asperity contact regions within the
macroasperity contact regions to reach the $N_{\rm cont}$ needed for wear particle formation 
even over a short sliding distance.

\begin{figure}
\includegraphics[width=0.47\textwidth,angle=0.0]{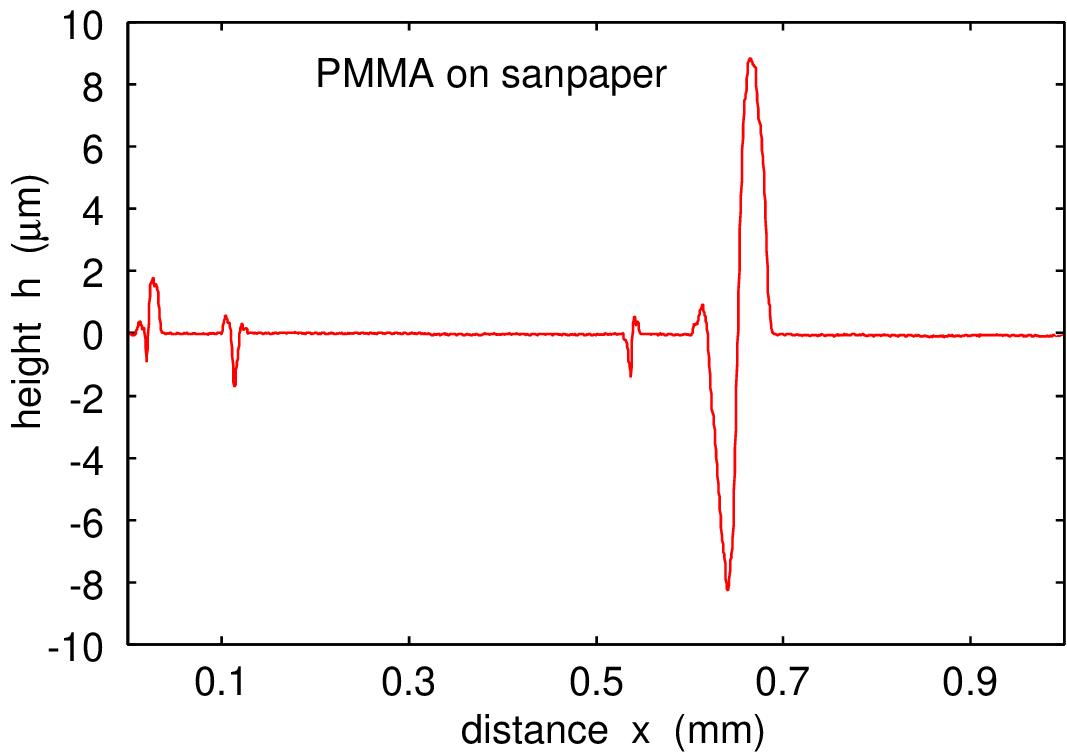}
\caption{\label{1x.2h.3.eps}
The height profile orthogonal to the ploughing tracks after a PMMA block was slid 
a short distance on a sandpaper P100 surface. 
}
\end{figure}

\vskip 0.3cm

{\bf 5 Role of plastic flow}

In the context of sliding wear, asperity contact regions may deform plastically at short length scales even when brittle fracture dominates at longer scales\cite{Moli1,Moli2,Moli}. 
This phenomenon has been observed even in very brittle materials such as silicon nitride. The tendency for plastic flow rather than crack propagation at small length scales can be understood based on the Griffith fracture criterion \cite{plast3,Rabi1,Rabi2,Moli1}: to remove a particle of linear size $r_0$, the local stress must be high enough that the stored elastic energy in a volume element $\sim r_0^3$ exceeds the fracture energy $\sim \gamma r_0^2$. This leads to the condition $\sigma > \sigma_{\rm c}$, where $\sigma_{\rm c}$ is defined by Eq. (4), for the removal of a particle. However, if $\sigma_{\rm c}$ exceeds the penetration hardness $\sigma_{\rm P}$ \textit{at the length scale} $r_0$, crack propagation cannot occur, and the material will instead flow plastically.

This principle is exploited in \textit{ductile mode cutting}, in which material is removed by plastic flow instead of brittle fracture, producing a smoother surface with minimal damage \cite{plast4,chip}. For example, when a hard asperity such as a diamond tip slides across the surface of a brittle solid at sufficiently low load (so that the contact region is very small), the material may be removed by cutting (producing micro or nano chips through plastic deformation) without the formation of surface cracks. This technique is commonly used to machine brittle materials such as silica glass, silicon, silicon nitride, or tungsten carbide \cite{plast4,chip}. Cutting in this context refers to the removal of material from the surface in the form of primary debris or microchips, with minimal lateral displacement, resembling conventional machining. A recent study has shown that a transition occurs from cutting dominated by shear deformation (plastic flow) to fracture-induced chip formation when the cutting depth exceeds a critical value \cite{PRL}.

Studies have shown that all earthquake fault surfaces exhibit anisotropic roughness, with grooves aligned along the fault sliding direction. Faults involve contact between the same type of material (usually granite), where asperities scratch the counter surface during slip, resulting in anisotropic roughness. However, Candela and Brodsky \cite{Can} have shown that below a critical length scale $\lambda^*$, the fault surface roughness becomes isotropic. They interpreted this as resulting from plastic deformation of the asperities at length scales $\lambda < \lambda^*$, while brittle fracture dominates at larger scales. Brodsky and coworkers have also suggested that studies of fault surface roughness can provide insights into the scale dependence of penetration hardness \cite{Brodsky, Brod}.

In many situations, plastic flow does not result in material removal but only in its displacement \cite{W1,W2}. 
This is particularly true when a plastically softer material slides on a harder solid.
When the slopes of asperities are not too steep and the sliding friction is small, ploughing tracks are formed. 
In such cases, the stress in the asperity contact regions is mainly compressive, which suppresses the formation of cracks, 
and the material is displaced to the sides of the grooves by plastic flow rather than being detached as wear particles, 
a process more likely to occur for sharper surface features. 
The amount of the displaced material can be estimated from line scan topography measurements on an initially flat surface of the test material. 
If sliding occurs only once and at low contact pressure, the spacing between ploughing tracks is relatively large. 
We define the flat regions between the tracks as the \textit{undeformed surface plane}. 
The volume of material removed as wear particles can be determined by comparing the material volumes below and above this undeformed surface plane.

To illustrate this, Fig. \ref{1x.2h.3.eps} shows a segment of a $10 \ {\rm mm}$ long line scan for PMMA that was slid a short distance over a sandpaper surface. An analysis of the ploughing tracks from the entire scan indicates that most 
of the material volume removed below the undeformed surface plane was displaced rather than lost as wear particles. 
Such material displacement by plastic flow occur to some extent also for a brittle material like silica glass (see Appendix A).

\begin{figure}
\includegraphics[width=0.47\textwidth,angle=0.0]{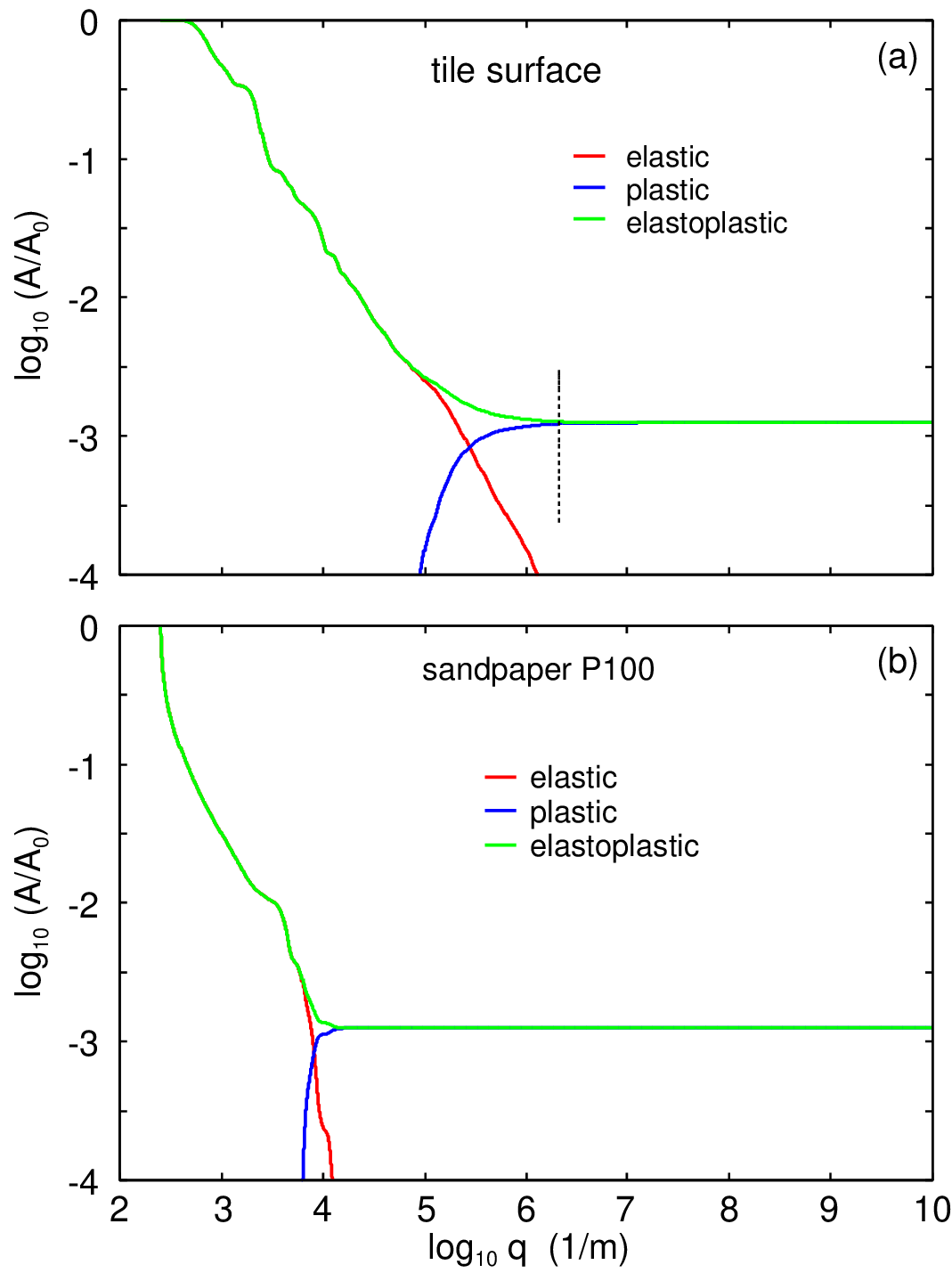}
\caption{\label{1logq.2Area.PMMA.plastic.eps}
The area of real contact as a function of the largest wavenumber $q$ included in the calculation, for PMMA on the tile surface (a) and on the sandpaper P100 surface (b). The wavenumber $q$ is related to the magnification $\zeta$ via $q = \zeta q_0$. The red and blue lines represent the elastic and plastic contact areas, respectively, and the green line indicates the elastic contact area for the elastoplastically deformed surface. The calculations assume elastoplastic contact with a Young's modulus of $E = 3 \ {\rm GPa}$, a Poisson ratio of $\nu = 0.3$, and a penetration hardness of $\sigma_{\rm P} = 0.4 \ {\rm GPa}$.}
\end{figure}

In Fig. \ref{1logq.2Area.PMMA.plastic.eps}(a), we show the area of real contact as a function of the cut-off wavenumber $q$ for PMMA on the tile surface. The cut-off wavenumber $q$ corresponds to the shortest wavelength roughness included in the calculation and is related to the magnification $\zeta$ through $q = \zeta q_0$. In the calculations, we have used a penetration hardness of $\sigma_{\rm P} = 0.4 \ {\rm GPa}$ for PMMA and assumed the tile surface to be rigid. When calculating the contact area for a given wavenumber $q = \zeta q_0$, only the long-wavelength roughness components with $q_0 < q < \zeta q_0$ are included. Note that all contact regions have yielded plastically when $q \approx 2 \times 10^6 \ {\rm m}^{-1}$, which corresponds to a wavelength of approximately $1 \ {\rm \mu m}$.

The sandpaper P100 surface exhibits a larger surface roughness power spectrum than the tile surface. For PMMA in contact with the sandpaper, plastic deformation begins at longer length scales, as shown in Fig. \ref{1logq.2Area.PMMA.plastic.eps}(b). For this surface, plastic yielding begins at $q \approx 10^4 \ {\rm m}^{-1}$, corresponding to a wavelength of approximately $0.3 \ {\rm mm}$, which is comparable to the average size of the sand particles. 

The steel surface is relatively smooth. In this case, plastic deformation occurs only at very short length scales, involving surface roughness components 
with wavelengths below approximately $100 \ {\rm nm}$. 
These components are not significant for the wear process considered in this study. Therefore, for the steel surface, PMMA is treated as a purely elastic material.

\vskip 0.3cm
{\bf 6 Comparing theory with experiments}

Here we compare the theoretical predictions for the wear rate with the experimental results for PMMA sliding on tile, sandpaper, and steel surfaces, as well as for silica glass sliding on the sandpaper surface.

\vskip 0.1cm
{\bf PMMA on tile, sandpaper, and steel}

We model PMMA as an elastoplastic material with a Young's modulus of $E = 3 \ {\rm GPa}$ and a Poisson ratio of $\nu = 0.3$. The penetration hardness of PMMA depends on the indentation time, as the deformation process is a stress-augmented, thermally activated flow. In this study, the indentation time is estimated as $\tau = r/v$, where $v$ is the sliding speed and $r$ is a characteristic length scale, approximately equal to the typical radius of a wear particle, $r \approx 3 \ {\rm \mu m}$. For a sliding speed of $v \approx 3 \ {\rm mm/s}$, the indentation time is estimated to be $\tau \approx 10^{-3} \ {\rm s}$.

For PMMA, the measured penetration hardness as a function of the strain rate is well described by the following empirical relation (in MPa) (see Ref. \cite{Bru}):
$$
\sigma_{\rm P} \approx 0.313 + 0.0325 \ {\rm log}_{10} \left( \tau_0/\tau \right), \eqno(20)
$$
where $\tau_0 = 1 \ {\rm s}$.

Using $\tau \approx 10^{-3} \ {\rm s}$, the penetration hardness is calculated as $\sigma_{\rm P} \approx 0.41 \ {\rm GPa}$.

The tile and steel surfaces have much higher elastic modulus and penetration hardness than PMMA and are therefore treated as rigid materials. The sandpaper consists of very hard and elastically stiff corundum (aluminum oxide) particles deposited on an elastically soft polymer film. As discussed in Sec. 3, this is accounted for by including only the substrate roughness components with wavenumber $q > 2\pi /D$, where $D$ is the average diameter of the corundum particles. For the steel and tile surfaces, we use the surface roughness power spectra shown in Fig. \ref{1logq.2logC.tile.steel.P100.eps}. For the sandpaper, we exclude the region with $q < 2\pi /D$. In all cases, the measured power spectra are linearly extrapolated to larger wavenumbers on a log-log scale. The slope of the extrapolated region corresponds to a Hurst exponent $H \approx 1$, but the exact form of this extrapolation is not critical for the wear rate calculations presented below.

To calculate the wear rate, we need the relation between $\Delta x$ and $\gamma$. This relationship has been experimentally studied for PMMA and varies slightly depending on the PMMA formulation\cite{Paris1}. The measured relation is well approximated by the following expression:
\begin{align*}
\Delta x &= 0, \quad {\rm for} \ \gamma < \gamma_0, \\
\Delta x &= a  \left (\sqrt{\gamma}-\sqrt{\gamma_0} \right )^2 \left (\frac{\sqrt{\gamma}-\sqrt{\gamma_0}}{\sqrt{\gamma_{\rm c}} - \sqrt{\gamma}} 
\right )^{\sqrt{\gamma_0/\gamma_{\rm c}}} \tag{21}
\end{align*}
for $\gamma_0 < \gamma < \gamma_{\rm c}$, where $\gamma_0 = 36.0 \ {\rm J/m^2}$, $\gamma_{\rm c} = 517.0 \ {\rm J/m^2}$, and $a = 5.3 \times 10^{-9} \ {\rm m^3/J}$. 
Note that $\Delta x \rightarrow \infty$  as $\gamma$ approaches $\gamma_{\rm c}$.

\begin{figure}
\includegraphics[width=0.47\textwidth,angle=0.0]{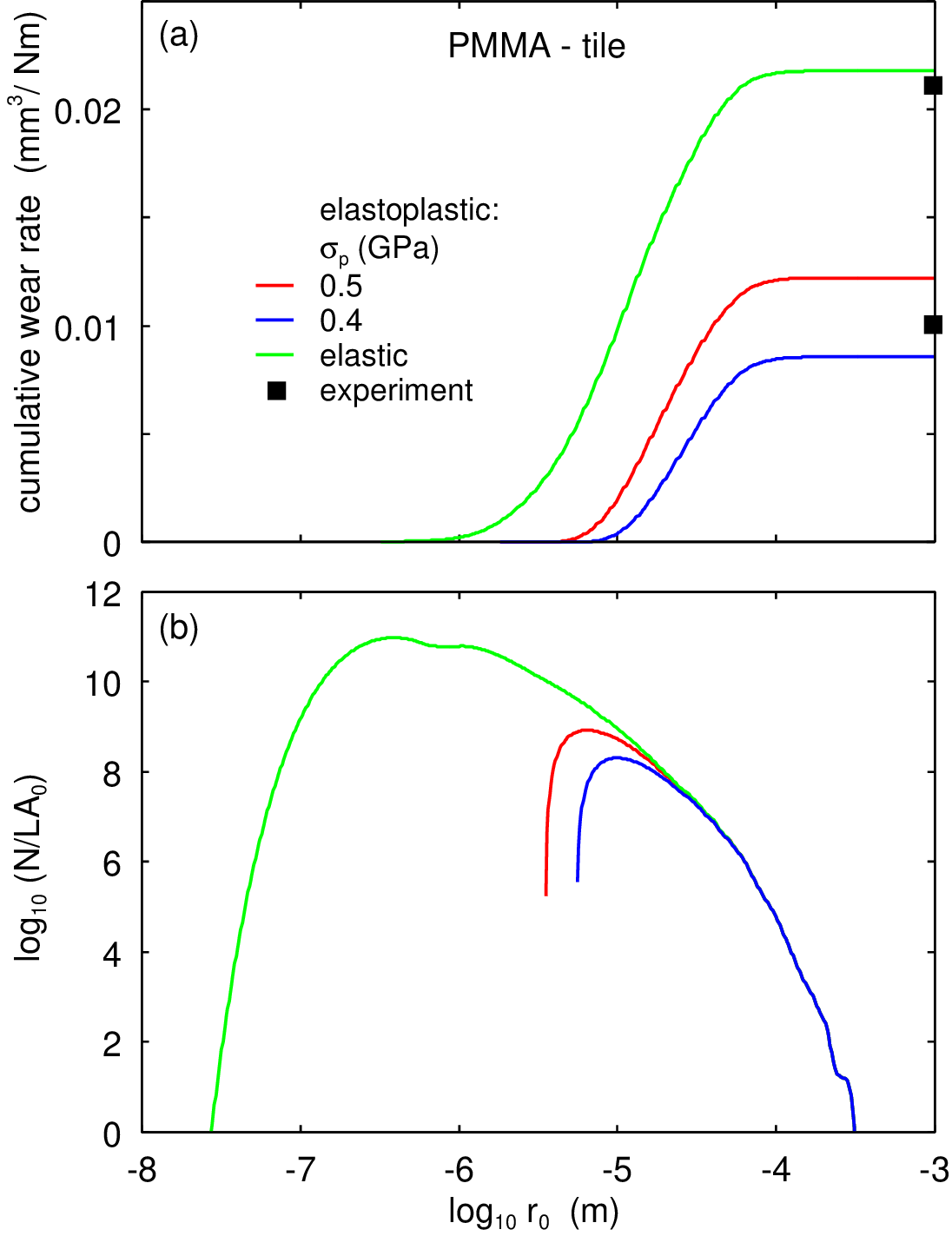}
\caption{\label{1logr0.2NoverL.PMMA.eps}
(a) The cumulative wear volume and (b) the number of generated particles as functions of the logarithm of the particle radius for PMMA sliding on a tile surface. The wear rates without plastic deformation and with plastic deformation (assuming $\sigma_{\rm P}=0.4 \ {\rm GPa}$, blue line) are approximately $0.22 \ {\rm mm^3/Nm}$ and $0.085 \ {\rm mm^3/Nm}$, respectively. The experimental wear rates are indicated by black squares ($0.021$ and $0.01 \ {\rm mm^3/Nm}$; see Fig. \ref{1distance.massloss.PMMA.tile.and.sandpaper.eps}). Calculations use $E=3 \ {\rm GPa}$, $\nu = 0.3$, and the measured relation between the crack-tip displacement $\Delta x (\gamma)$ and the tearing energy $\gamma$, shown in Fig. \ref{1logGamma.2Dx.eps}(a). The friction coefficient is $\mu = 0.5$, the nominal contact area $A_0 = 0.002 \ {\rm m^2}$, and the nominal contact pressure $\sigma_0 = 0.052 \ {\rm MPa}$, as in the experiment described in Sec. 3.
}
\end{figure}

Using the power spectra shown in Fig. \ref{1logq.2logC.tile.steel.P100.eps} and the relation between $\Delta x$ and $\gamma$ given by equation (21), we present in Fig. \ref{1logr0.2NoverL.PMMA.eps}(a) the cumulative wear volume and in Fig. \ref{1logr0.2NoverL.PMMA.eps}(b) the number of generated particles as functions of the logarithm of the particle radius for the PMMA-tile system. The green lines represent the results obtained without considering plastic deformation using equation (5), with the power spectrum indicated by the red line in Fig. \ref{1logq.2Area.PMMA.plastic.eps}. The red and blue lines correspond to calculations that include plasticity using equation (7).

The calculated wear rate for $\sigma_{\rm P} = 0.4 \ {\rm GPa}$ [blue line in Fig. \ref{1logr0.2NoverL.PMMA.eps}(a)] is $\Delta V/LF_{\rm N} \approx 0.0085 \ {\rm mm^3/Nm}$, which is consistent with the experimental values ($0.021$ and $0.010$ from two separate measurements; see Sec. 3). The peak in the number of generated particles in Fig. \ref{1logr0.2NoverL.PMMA.eps}(b) occurs at particle sizes approximately one order of magnitude smaller than in previous studies of rubber wear. The optical method we used does not give sufficient contrast to accurately measure the PMMA particle sizes experimentally. The number of generated particles in different size ranges can be determined from Fig. \ref{1logr0.2NoverL.PMMA.eps}(b) using the two-interval separation method described previously.

\begin{figure}
\includegraphics[width=0.47\textwidth,angle=0.0]{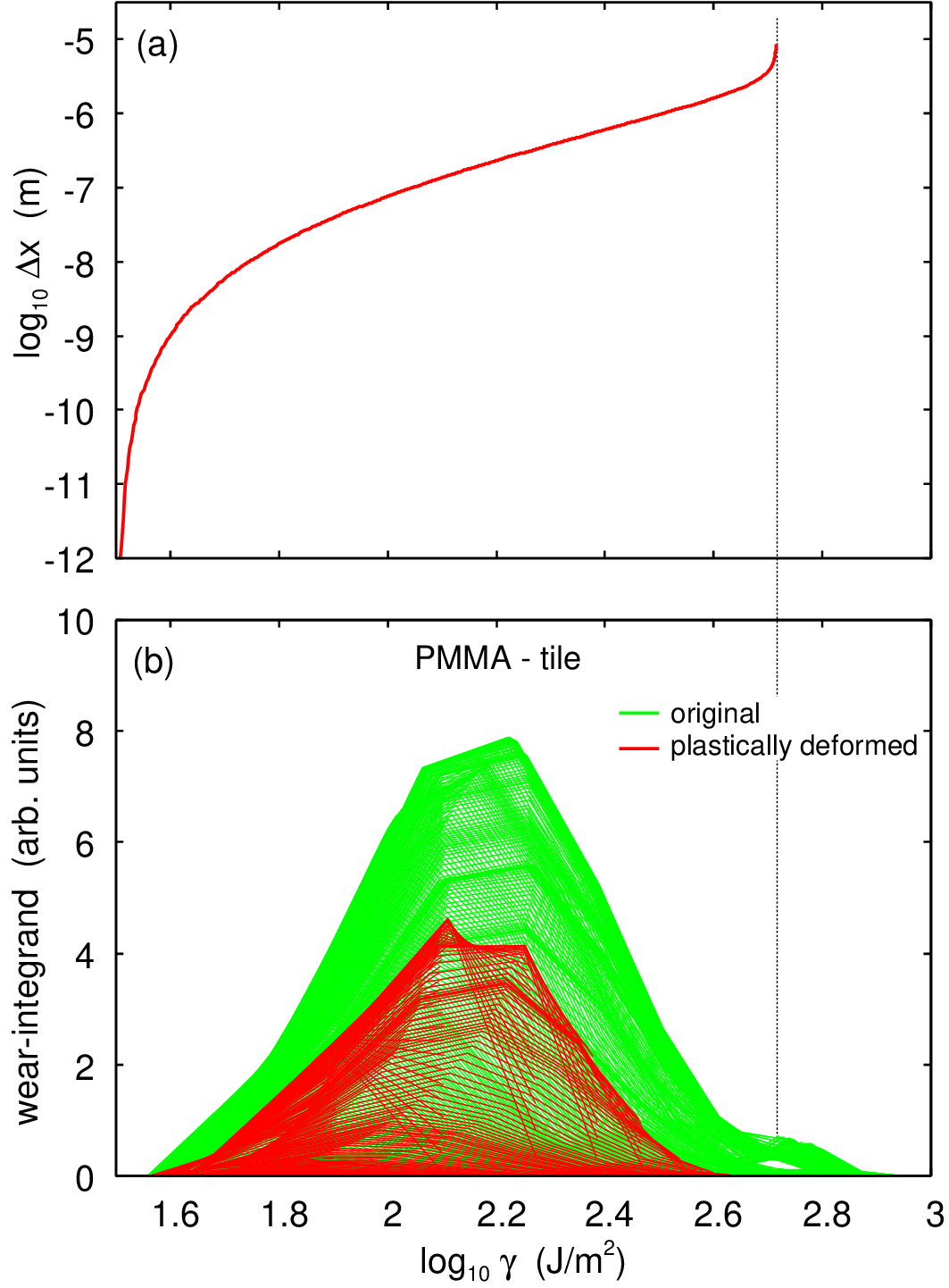}
\caption{\label{1logGamma.2Dx.eps}
(a) Relationship between the crack tip displacement per oscillation and the tearing (or crack) energy $\gamma$ for PMMA, based on the measurements presented in Ref. \cite{Paris1}. 
(b) Integrand in equation (15) as a function of $\gamma$ for all magnifications (or particle radii $r_0$), assuming elastic contact without plasticity (green lines) and elastoplastic contact with $\sigma_{\rm P} = 0.4 \ {\rm GPa}$ (red lines). 
}
\end{figure}

The relationship between $\gamma$ and $\Delta x$ depends on the specific type of PMMA, but it generally follows the form given in equation (21), which is also depicted in Fig. \ref{1logGamma.2Dx.eps}(a). Note that $\Delta x$ diverges as $\gamma$ approaches the critical value $\gamma_{\rm c}$. In Fig. \ref{1logGamma.2Dx.eps}(b), we display the integrand of equation (15) for PMMA on the tile surface as a function of $\gamma$ for all magnifications (or particle radii $r_0$), assuming no plastic flow (green lines) and including plasticity (red lines). Although the integration variable in equation (15) is pressure, each pressure value corresponds to a tearing energy as defined by equation (4). The red and green areas represent the superposition of many curves corresponding to various magnifications or particle radii.

\begin{figure}
\includegraphics[width=0.47\textwidth,angle=0.0]{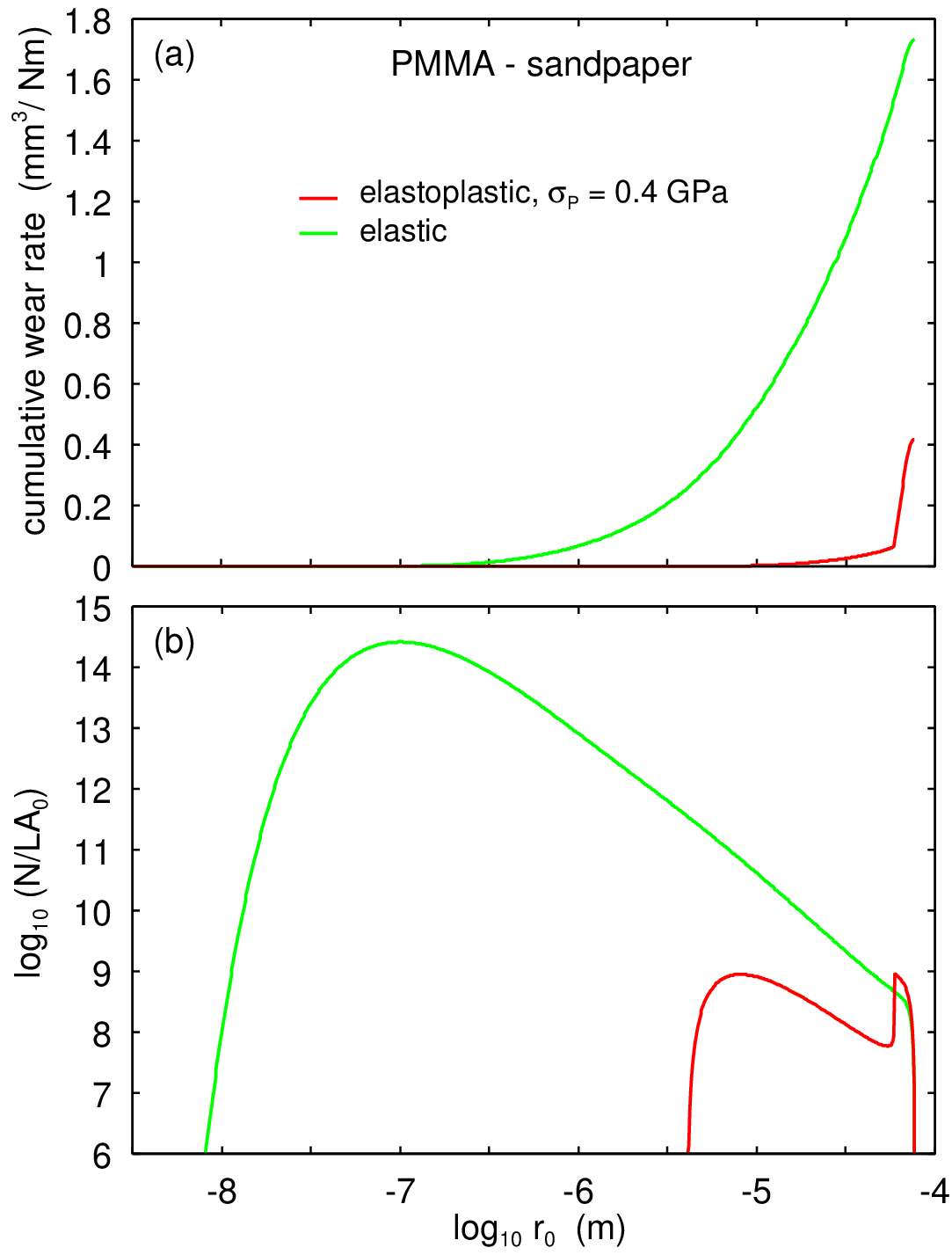}
\caption{\label{1logr0.2wearCumulative.PMMA.P100cut.eps}
(a) Cumulative wear volume and (b) number of generated particles as a function of the logarithm of the particle radius for PMMA on sandpaper P100.
The wear rates without and with plastic deformation (assuming $\sigma_{\rm P} = 0.4 \ {\rm GPa}$) are approximately $1.74 \ {\rm mm^3/Nm}$ and $0.40 \ {\rm mm^3/Nm}$, respectively.
The measured wear rate is approximately $0.39 \ {\rm mm^3/Nm}$ (see Fig. \ref{1distance.massloss.PMMA.tile.and.sandpaper.eps}).
}
\end{figure}

Fig. \ref{1logr0.2wearCumulative.PMMA.P100cut.eps}(a) shows the cumulative wear volume and (b) the number of generated particles as functions of the logarithm of the particle radius for PMMA sliding on sandpaper. We have used the power spectrum of the sandpaper surface for $q > 2\pi /D$ and the measured friction coefficient $\mu = 0.60$. The green line is calculated using equation (5) without plasticity, and the red lines include plastic flow using equation (7). The wear rates without and with plastic deformation (with $\sigma_{\rm P} = 0.4 \ {\rm GPa}$) are approximately $1.74 \ {\rm mm^3/Nm}$ and $0.40 \ {\rm mm^3/Nm}$, respectively, while the measured wear rate is approximately $0.39 \ {\rm mm^3/Nm}$ 
(see blue line in Fig. \ref{1distance.massloss.PMMA.tile.and.sandpaper.eps}).

\begin{figure}
\includegraphics[width=0.47\textwidth,angle=0.0]{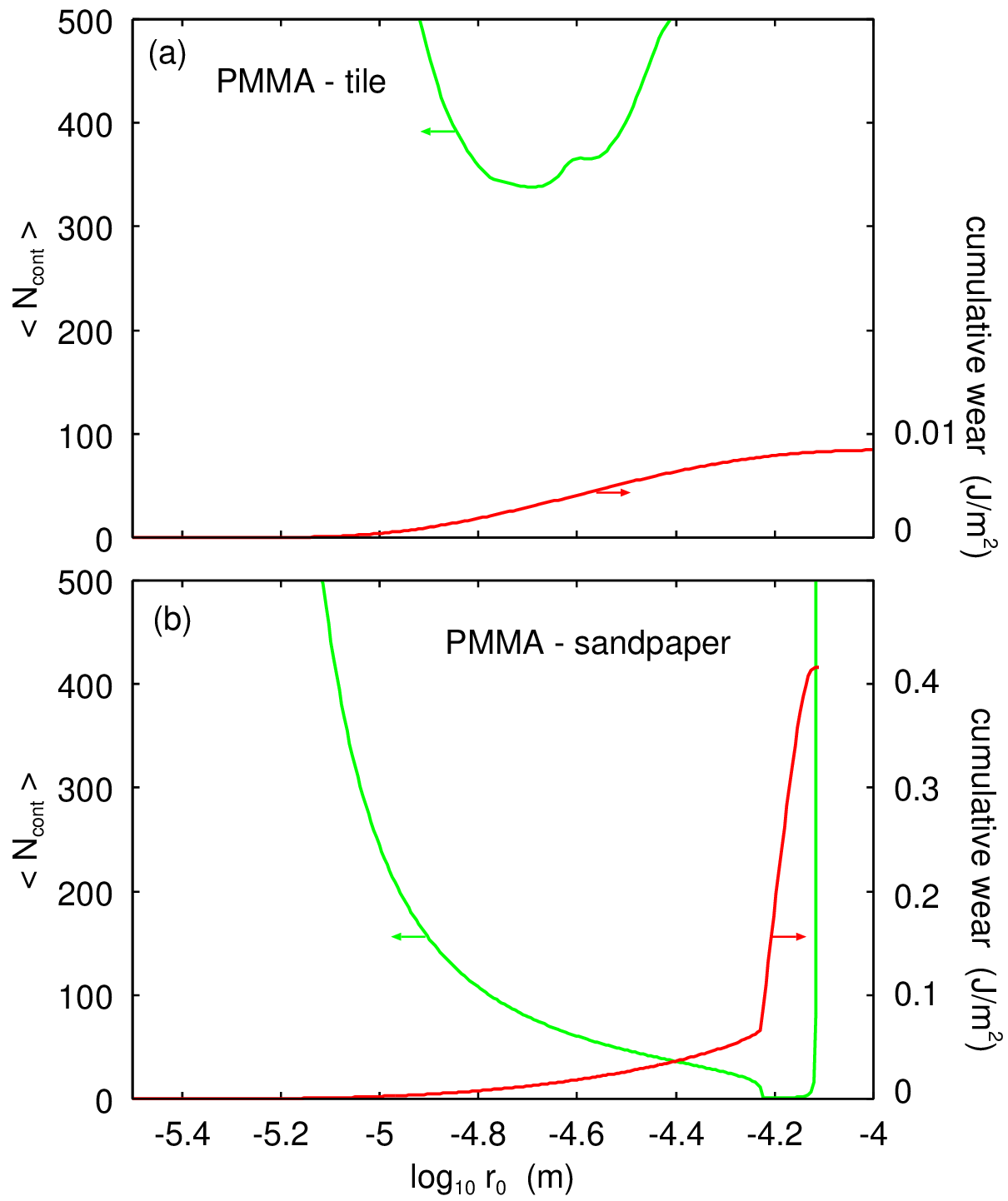}
\caption{\label{1logr0.2Ncont.PMMA.tile.new1.eps}
Cumulative wear volume (red line) and the effective number of asperity contacts $\langle N_{\rm cont} \rangle$ required to remove a wear particle (green line) as functions of the logarithm of the wear particle radius.
Results are shown for elastoplastic contact (assuming $\sigma_{\rm P} = 0.4 \ {\rm GPa}$) for PMMA on (a) the tile surface and (b) the sandpaper surface.
}
\end{figure}

For PMMA on the tile surface, several hundred contacts with wear asperities are required to remove a single PMMA wear particle. This is illustrated in Fig. \ref{1logr0.2Ncont.PMMA.tile.new1.eps}(a), where the green line shows the effective number of contacts $\langle N_{\rm cont} \rangle$ needed to detach a wear particle as a function of the logarithm of the wear particle radius. The red line in the figure, also shown in Fig. \ref{1logr0.2NoverL.PMMA.eps}(a), represents the cumulative wear volume, including plastic deformation, as a function of the logarithm of the wear particle radius. Approximately 70 percent of the wear mass is attributed to particles removed in fewer than $\sim 500$ contacts with wear asperities.

For PMMA on the sandpaper surface, most wear particles are removed in a single contact between the PMMA and the corundum wear asperities. This is demonstrated in Fig. \ref{1logr0.2Ncont.PMMA.tile.new1.eps}(b), where the sharp increase in cumulative wear volume occurs when $\langle N_{\rm cont} \rangle \approx 0$, corresponding to the condition $\Delta x \gg r_0$.

Using the same parameters as above but with $\mu = 0.2$, for PMMA on the polished steel surface, and with the surface power spectrum shown in Fig. \ref{1logq.2logC.tile.steel.P100.eps} (blue line), the predicted wear rate is $\Delta V/LF_{\rm N} \approx 10^{-21} \ {\rm mm^3/Nm}$. This indicates negligible wear contribution from the considered mechanism. This prediction is consistent with our experimental findings, where no measurable wear was detected after sliding over a distance of $4 \ {\rm m}$ on the steel surface. Given the resolution of our balance ($0.1 \ {\rm mg}$), this corresponds to an experimental upper limit for the wear rate of approximately $10^{-4} \ {\rm mm^3/Nm}$.

\begin{figure}
\includegraphics[width=0.47\textwidth,angle=0.0]{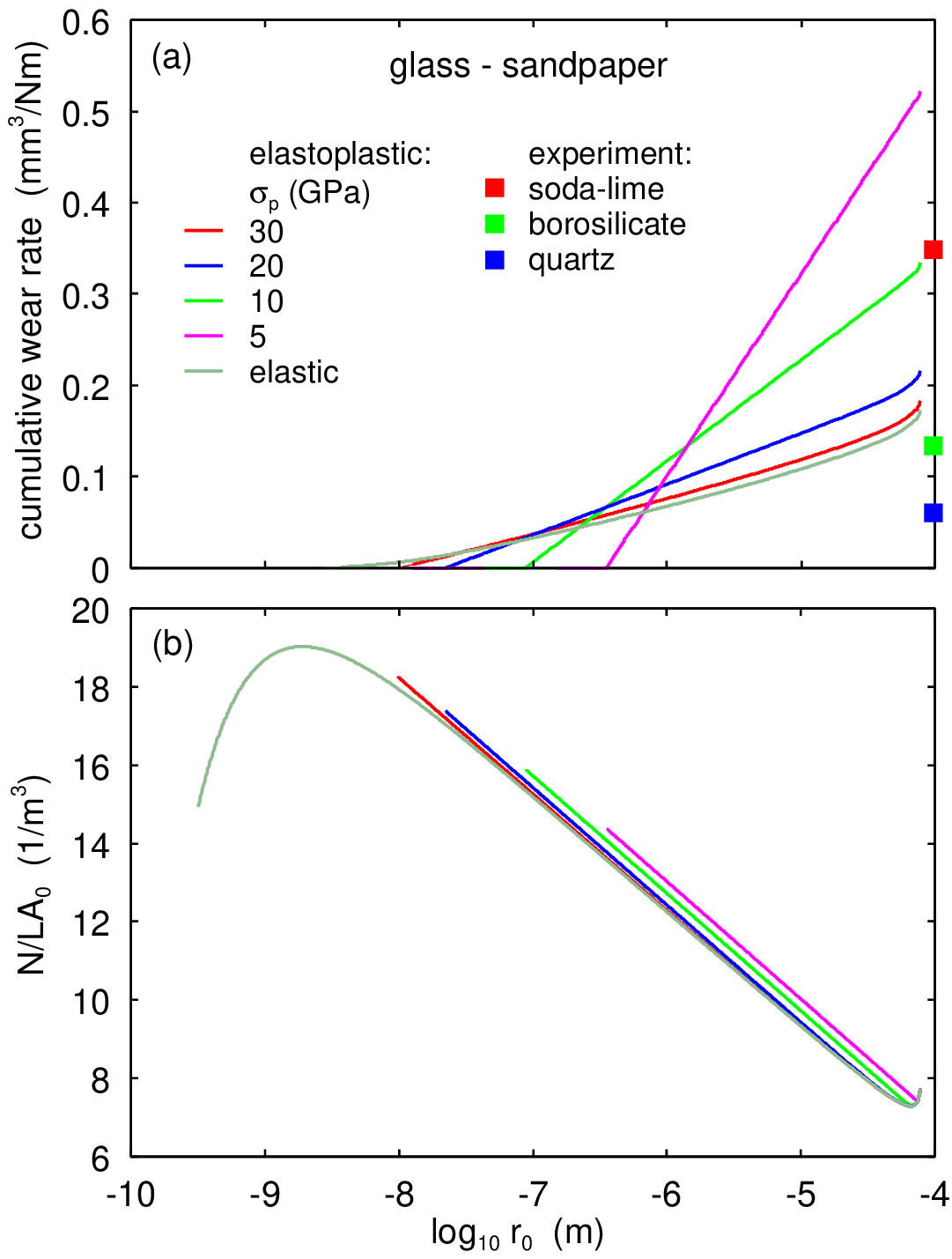}
\caption{\label{1logR0.2logWearRate.eps}
(a) Cumulative wear volume and (b) number of wear particles as functions of the logarithm of the wear particle radius for glass sliding on sandpaper P100. Results are shown for different values of penetration hardness. The curve labeled ``elastic'' corresponds to infinite hardness, which implies no plastic deformation. The nominal contact area is $A_0 = 0.002 \ {\rm m^2}$ and the nominal contact pressure is $\sigma_0 = 0.052 \ {\rm MPa}$, as in the experiment described in Sec. 3. Calculations use $E = 70 \ {\rm GPa}$, $\nu = 0.3$, and a friction coefficient of $\mu = 0.34$.
}
\end{figure}

\vskip 0.1cm
{\bf Silica-lime, borosilicate and quartz glass on sandpaper}

In Sec. 3, we studied the sliding wear of window glass (soda-lime), borosilicate glass, and quartz (crystalline ${\rm SiO_2}$) on sandpaper. The sandpaper consists of corundum particles (crystalline ${\rm Al_2O_3}$), which are both elastically stiffer and plastically harder than the glass materials, and are treated as rigid in our calculations. For silica glass and quartz, the Young's modulus is approximately $E \approx 70 \ {\rm GPa}$, while for corundum it is approximately $E \approx 350 \ {\rm GPa}$. The penetration hardness of corundum is $\sigma_{\rm P} \approx 22 \ {\rm GPa}$, which is higher than that of quartz, $\sigma_{\rm P} \approx 12 \ {\rm GPa}$ (see Ref. \cite{quartz}), borosilicate glass, $\sigma_{\rm P} \approx 8 \ {\rm GPa}$ (see Ref. \cite{boro}), and soda-lime glass, $\sigma_{\rm P} \approx 6{-}11 \ {\rm GPa}$ (see Ref. \cite{lime}), depending on the loading rate.

Fig. \ref{1logR0.2logWearRate.eps}(a) shows the calculated cumulative wear volume, and Fig. \ref{1logR0.2logWearRate.eps}(b) shows the number of wear particles, both as functions of the logarithm of the wear particle radius, for glass surfaces with different penetration hardness values sliding on sandpaper P100. The curve labeled ``elastic'' corresponds to infinite hardness, which implies no plastic deformation. The calculated wear rate for $\sigma_{\rm P} = 10 \ {\rm GPa}$ is $\Delta V/LF_{\rm N} \approx 0.33 \ {\rm mm^3/Nm}$, which is in close agreement with the experimentally measured wear rate for window glass ($0.34 \ {\rm mm^3/Nm}$; see Fig. \ref{1distance.2massloss.allGLASS.eps}). 

However, the current theoretical model does not account for the observed reduction in wear rates for borosilicate glass and quartz, which are approximately $0.38$ and $0.16$ times smaller than the wear rate of window glass, respectively (see again Fig. \ref{1distance.2massloss.allGLASS.eps}).

\begin{figure}
\includegraphics[width=0.47\textwidth,angle=0.0]{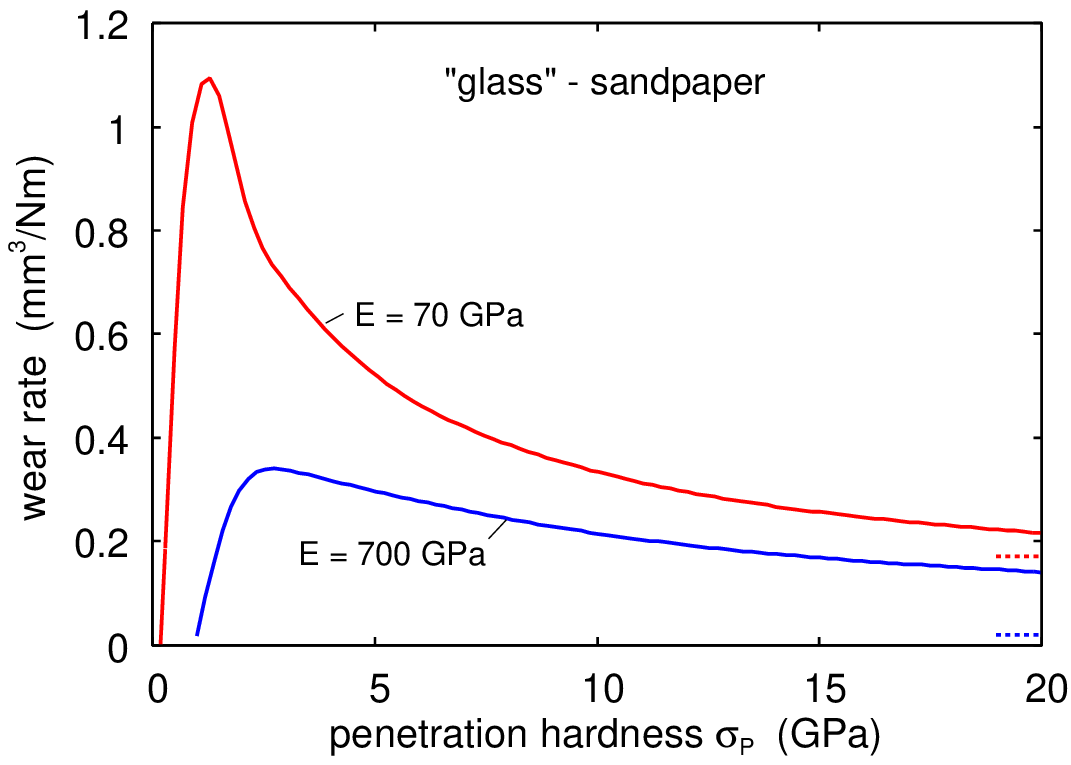}
\caption{\label{1sigmaP.2wearate.twoE.1.eps}
The wear rate $\Delta V/F_{\rm N}L$ (in ${\rm mm^3/Nm}$) as a function of the penetration hardness for glass sliding on the sandpaper P100 surface. The red and blue curves correspond to Young's modulus $E = 70 \ {\rm GPa}$ and $E = 700 \ {\rm GPa}$, respectively. In both cases, the Poisson ratio is $\nu = 0.3$ and the friction coefficient is $\mu = 0.34$. The dotted red and blue lines indicate the wear rates predicted under the assumption of purely elastic contact ($\sigma_{\rm P} = \infty$).
}
\end{figure}

\vskip 0.3cm
{\bf 7 Discussion}

Many wear equations have been developed, but the most widely used is probably the Archard wear equation: 
$${\Delta V \over LA_0} = K {p_0 \over \sigma_{\rm P}} \eqno(22)$$
where $p_0$ is the nominal contact pressure (assumed constant), and $A_0$ is the nominal contact area,
related to the applied normal force via $F_{\rm N} = p_0 A_0$. 
Equation (22) can also be written as
$${\Delta V \over F_{\rm N} L} = {K\over \sigma_{\rm P}} \eqno(23)$$
The parameter $K$ is dimensionless but depends on the specific system under investigation, and it has been found to span a wide range of values. 

Equation (23) assumes that all contact regions, when observed at the highest (atomic) resolution,
have undergone plastic yielding, and that wear results from the removal of fragments of material from the area of real contact. 
However, this relation (with $K > 0$) cannot hold universally. If the elastic energy $U_{\rm el}$ stored in the contact regions is less than the critical energy $U_{\rm cr}$ on all relevant length scales, then no wear particles will form, even if the material has yielded plastically.

Furthermore, even when wear does occur, Eq. (22) is generally not valid, even under the condition that all contact regions yield plastically. 
This point is illustrated in Fig. \ref{1sigmaP.2wearate.twoE.1.eps} for a hypothetical glass-sandpaper system.

Fig. \ref{1sigmaP.2wearate.twoE.1.eps} shows the calculated wear rate $\Delta V/F_{\rm N}L$ (in ${\rm mm^3/Nm}$) as a function of the penetration hardness $\sigma_{\rm P}$ for a glass surface sliding on the sandpaper P100 surface. The red and blue curves correspond to Young's modulus $E = 70 \ {\rm GPa}$ and $E = 700 \ {\rm GPa}$, respectively, with $\nu = 0.3$ and $\mu = 0.34$. The dotted red and blue lines represent the wear rate under the assumption of purely elastic contact ($\sigma_{\rm P} = \infty$).

For $\sigma_{\rm P} > 2.5 \ {\rm GPa}$ the wear rate decreases with increasing $\sigma_{\rm P}$. This results from the reduction in the area of real contact as $\sigma_{\rm P}$ 
increases. A decrease in the wear rate with increasing $\sigma_{\rm P}$ is also predicted by the Archard equation. However, we observe a weaker $\sigma_{\rm P}$ dependency 
than the $1/\sigma_{\rm P}$ dependency predicted by (22). 
This is due to the increase in the stress in the contact region as $\sigma_{\rm P}$ increases.
As $\sigma_{\rm P}$ decreases below a critical value (here $\approx 2.5 \ {\rm GPa}$), the wear rate decreases toward zero. 
This is expected because, at low $\sigma_{\rm P}$, the interfacial stress is everywhere too small to generate wear particles. 
That is, the condition $U_{\rm el} < U_{\rm cr}$ is satisfied across all relevant length scales. 
This non-monotonic behavior contrasts with traditional models such as Archard's law, which predict a monotonic decrease in wear with increasing hardness.

Both crystalline and amorphous solids exhibit penetration hardness that depends on temperature, deformation rate, and the indentation size. Plastic deformation in crystalline solids involves the motion of dislocations, while in amorphous solids it involves atomic rearrangements within small volume elements, typically a few nanometers in size. The motion or generation of dislocations, as well as atomic rearrangements in amorphous materials, requires overcoming energy barriers (for dislocations, these are related to the so-called Peierls stress). 

Thus, at nonzero temperatures, plastic flow in both crystalline and amorphous solids involves stress-augmented thermally activated processes. As a result, the penetration hardness decreases with increasing temperature or decreasing deformation rate.

During sliding, the asperity contact time is on the order of $\tau = r_0/v$, where $r_0$ is the width of the contact region. The corresponding frequency is $\omega = 1/\tau = v/r_0$, which is typically much higher than the frequencies used in indentation studies. For example, if $v = 1 \ {\rm m/s}$ and $r_0 = 10 \ {\rm \mu m}$, then $\omega = 10^5 \ {\rm s}^{-1}$, while most indentation experiments are typically performed on the time scale of seconds. At such high deformation frequencies, the influence of temperature decreases, and the penetration hardness increases. Therefore, the penetration hardness relevant for wear calculations may be higher than the values listed in standard data tables. This effect could contribute to discrepancies between experimental results and theoretical predictions for quartz.

Furthermore, experiments have shown that the hardness of both crystalline and amorphous solids increases with decreasing indentation size. For quartz, this size effect has been studied in Ref. \cite{lime}, where nanoindentation at room temperature yielded a penetration hardness of approximately $17.5 \pm 2 \ {\rm GPa}$ at strain rates of $10^{-2}$ to $10^{-1} \ {\rm s}^{-1}$. The same study also showed that the penetration hardness decreases nearly linearly (by a factor of 2) as the temperature increases from $T = 20^\circ {\rm C}$ to $T = 500^\circ {\rm C}$. These observations indicate that penetration hardness is governed by stress-augmented thermally activated processes even at room temperature, and therefore may be higher at larger strain rates. If the penetration hardness of quartz is significantly higher than that of soda-lime glass, it could partly explain the larger wear rate observed for the latter.

We note that the Persson contact mechanics theory can be extended to account for size-dependent hardness \cite{Preview,Brodsky}. A numerical implementation of this extension was presented in an important study by Lambert and Brodsky \cite{Brodsky}.

In the study presented above, it was found that for PMMA sliding on the tile surface, the effective number of contacts $N_{\rm cont}$ required to detach a wear particle is much larger than unity. However, it is not clear whether, during relatively short sliding distances, points on the PMMA surface actually undergo contact with wear asperities as many times as expected from the large value of $N_{\rm cont}$. This observation suggests that the relationship $\Delta x (\gamma)$ may need to be interpreted within a probabilistic framework, which will be discussed in the next section.

The Paris equation provides the crack-tip displacement $\Delta x$ as a function of the tearing energy $\gamma$ when a crack is subjected to oscillating strain or stress, such as those arising from external boundary forces acting on the solid. When $\gamma$ is close to the fatigue limit $\gamma_0$, a large number of oscillation cycles is required before the crack-tip displacement becomes measurable. It is commonly assumed that cracks propagate continuously with the number of stress cycles. However, in this study, we propose an alternative scenario.

We assume that when $\gamma < \gamma_{\rm c}$, crack propagation occurs via stress-assisted thermally activated bond-breaking events. In this process, segments at the crack tip move in temporally irregular and discrete steps, each of which may be significantly larger than an atomic spacing. Therefore, if $U_{\rm el} > U_{\rm cr}$, the factor $1/(1 + r_0/\Delta x)$ may represent the probability of forming a wear particle during a single contact event of size approximately $r_0$, rather than the inverse of the number of contacts required to form such a particle. For example, if $\Delta x = 10^{-11} \ {\rm m}$ per cycle (or less), which is expected when $\gamma$ is sufficiently close to $\gamma_0$, then during most stress cycles no significant crack advancement occurs. However, occasionally, a segment along the crack front may displace by a characteristic distance that exceeds atomic dimensions.

The wear process described above results from elastic energy that is temporarily stored in asperity contact regions during sliding. If this elastic energy on a given length scale $r_0$ exceeds the critical threshold $U_{\rm el}>U_{\rm cr}$, crack propagation may remove particles or fragments of size $r_0$ from the sliding surfaces. On the other hand, if the surface roughness is sufficiently small such that $U_{\rm el} < U_{\rm cr}$ across all asperity contact regions and on all relevant length scales, this wear mechanism will not occur.

This behavior is demonstrated in the case of PMMA sliding on a polished steel surface, where the predicted wear rate is $\Delta V /LF_{\rm N} \approx 10^{-21} \ {\rm mm^3/Nm}$. For comparison, on a tile surface with approximately ten times higher root-mean-square roughness ($\approx 3 \ {\rm \mu m}$ compared to $0.3 \ {\rm \mu m}$ for the steel surface), the predicted wear rate is $\Delta V /LF_{\rm N} \approx 0.01 \ {\rm mm^3/Nm}$. In general, there exists an abrupt transition in wear behavior: as surface roughness decreases, the system transitions from a regime of relatively high wear rates to one of extremely low wear.
For elastic contact, the abruptness depends on the fact that the stress probability distribution in asperity contact regions decreases rapidly with increasing stress, as described by (5).

\begin{figure}
\includegraphics[width=0.47\textwidth,angle=0.0]{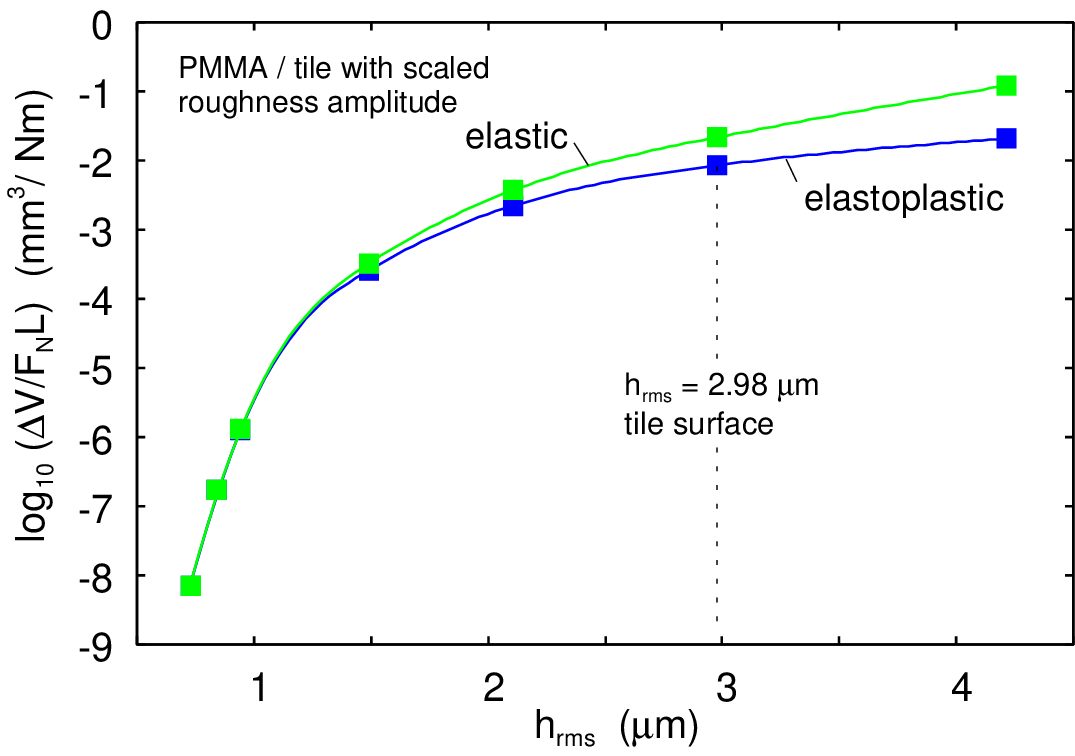}
\caption{\label{1hrms.2logWearRate.ElasticPlastic.eps}
The logarithm of the wear rate as a function of the root-mean-square roughness $h_{\rm rms}$ for PMMA in contact with a tile surface. The surface roughness is scaled to obtain different roughness amplitudes while maintaining the same fractal characteristics. The blue line represents the result for elastoplastic contact with a penetration hardness $\sigma_{\rm P} = 0.4 \ {\rm GPa}$, and the green line corresponds to purely elastic contact (no plastic deformation).
}
\end{figure}

\begin{figure}
\includegraphics[width=0.47\textwidth,angle=0.0]{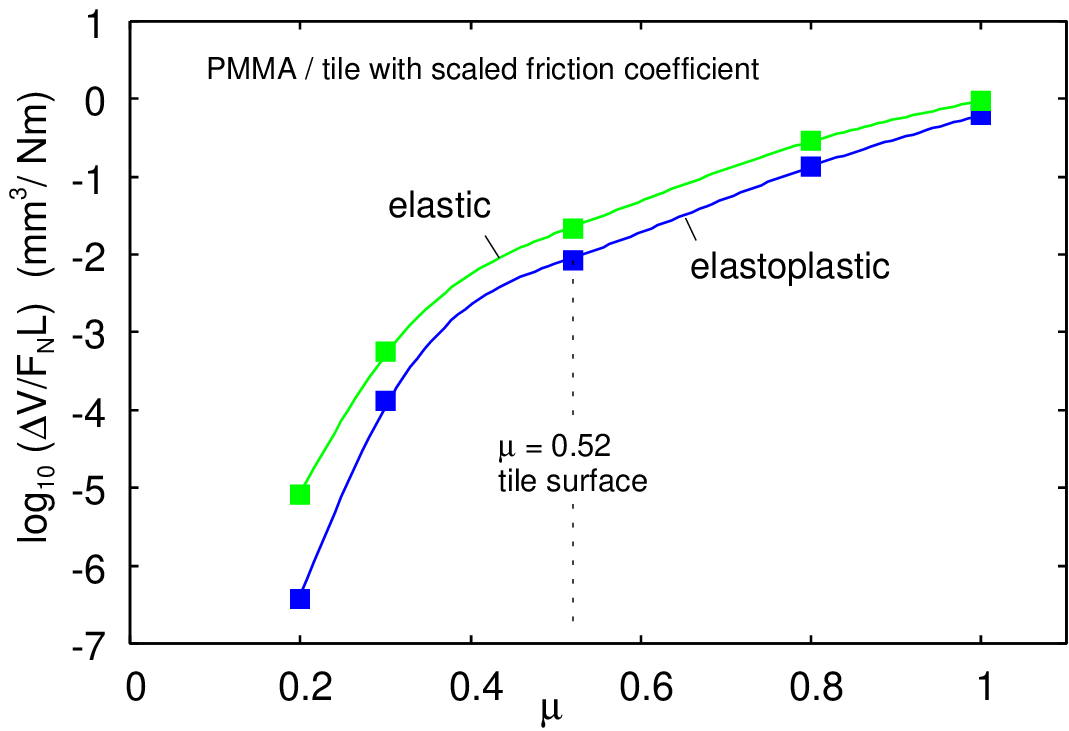}
\caption{\label{1mu.2logWearRate.elast.plast.eps}
The logarithm of the wear rate as a function of the friction coefficient for PMMA in contact with a tile surface. The friction coefficient is varied from the experimentally measured value of $0.52$ to both lower and higher values. The blue line shows the result for elastoplastic contact with a penetration hardness $\sigma_{\rm P} = 0.4 \ {\rm GPa}$, and the green line shows the result for elastic contact (without plastic deformation).
}
\end{figure}

This transition is illustrated in Fig. \ref{1hrms.2logWearRate.ElasticPlastic.eps}, which shows the logarithm of the wear rate as a function of the root-mean-square roughness $h_{\rm rms}$, obtained by scaling the power spectrum of the tile surface. For smooth surfaces with small $h_{\rm rms}$, no plastic flow occurs, and the elastoplastic results (blue line) are identical to those obtained under purely elastic conditions (green line). Similarly, a decrease in the friction coefficient reduces $U_{\rm el}$. For the PMMA and tile system, numerical calculations show that when $\mu$ falls below approximately $0.15$, the wear rate vanishes, as illustrated in Fig. \ref{1mu.2logWearRate.elast.plast.eps}.

Many studies have been conducted on UHMWPE sliding against very smooth counter surfaces, as these systems are of significant interest for artificial joint applications. The generation of wear debris has been identified as a major cause of failure in total joint replacements. In most of these systems, one of the bearing surfaces is made of a hard, extremely smooth metal or ceramic material, while the other surface consists of UHMWPE. Wear particles generated from UHMWPE during sliding can be released into the surrounding tissues, where they cause adverse cellular responses, potentially leading to bone resorption and implant loosening. Therefore, reducing the volume and number of UHMWPE wear particles is essential for improving the long-term clinical performance of total artificial joints.

In typical artificial joint systems, the counter surface in contact with the UHMWPE has an rms roughness below $0.03 \ {\rm \mu m}$. For such smooth surfaces, the wear mechanism described earlier cannot occur, and the observed wear rate is typically very low, on the order of $10^{-8} \ {\rm mm^3/Nm}$. The most probable origin of this wear is polymer asperity wear. The UHMWPE surfaces in these systems often exhibit surface roughness with amplitudes on the order of micrometers (rms roughness of approximately $1 \ {\rm \mu m}$). Under load, many of these asperities undergo elastoplastic deformation. During sliding, tensile stresses can develop at the base of the asperities, particularly on their trailing edges, which may lead to crack propagation or stress corrosion and the formation of wear particles. This mechanism likely dominates wear in artificial joints and in other cases where polymers slide against very smooth surfaces.

The wear process discussed above is not the only possible mechanism but is expected to be dominant unless the counter surface roughness is extremely small. In such cases, significantly lower wear rates are expected compared to those for rougher surfaces such as tile or sandpaper.

Another important wear mechanism is adhesive wear, which is particularly relevant for metals \cite{metal}. In asperity contact regions, local pressures can exceed the strength of the native oxide layer present on most metals, leading to direct metal-to-metal contact. If the materials are similar in composition, such as steel sliding on steel, cold-welded junctions can form and result in material transfer. This phenomenon has been confirmed in experiments using radioactive tracers to monitor material exchange.

The transferred material typically includes fragments of the oxide layer, and the exposed fresh metal quickly oxidizes, forming a new transferred film enriched with oxides. This film is often more weakly bound to the substrate than the bulk metal, and after sufficient sliding, oxide-rich particles may be generated. These particles may initially remain trapped between the surfaces, but over time they detach and contribute to the wear debris.

\vskip 0.3cm
{\bf 8 Summary and conclusion}

In this study, we extended a previously developed elastic wear model based on fatigue crack propagation to account for elastoplastic deformation. 
The new model was applied to predict wear rates of PMMA and glass materials sliding under dry conditions. In particular, for PMMA and soda-lime glass, 
the model shows good agreement with the experimental measurements, significantly improving upon predictions based solely on elastic behavior.

For PMMA, plasticity plays a significant role. However, even for silica glass, we observe an increase in the wear rate by a factor of $\sim 2$ due to plasticity, assuming a penetration hardness of about $10 \ {\rm GPa}$. If the penetration hardness were instead $20 \ {\rm GPa}$, the predicted influence of plasticity on the wear rate would be negligible. Since the penetration hardness may increase at short length scales and with increasing deformation rate, it is not clear how important plasticity is for quartz under actual wear conditions.

A key finding of this work is that the wear rate exhibits a non-monotonic dependence on the penetration hardness. Specifically, for relatively soft materials, 
the wear rate increases with hardness, while for harder materials, it decreases. This behavior deviates from the monotonic trends predicted by traditional 
wear models such as Archard's law and highlights the importance of incorporating material-specific deformation mechanisms into wear modeling.

We presented experimental data for the wear rate of PMMA sliding on tile, sandpaper, and polished steel surfaces, as well as for soda-lime, borosilicate, and quartz glass sliding on sandpaper. The theoretical predictions from the extended elastoplastic model were compared with the measured results. While the overall agreement is satisfactory, two unresolved issues remain:

(a) The theory predicts similar wear rates for window (soda-lime-silica) glass and quartz (crystalline ${\rm SiO_2}$), as both materials exhibit nearly identical elastic modulus and similar 
penetration hardness. Experimentally, however, the wear rate for quartz is approximately $0.16$ times that of window glass. This discrepancy may be attributed to the 
structural differences between amorphous and crystalline materials. In crystalline solids such as quartz, plasticity occurs primarily through the generation and motion of dislocations, 
whereas in amorphous materials like window glass, plastic deformation is governed by local atomic rearrangements within nanometer-sized domains. 
These distinct deformation mechanisms may result in differences in work-hardening behavior and flow properties, which are not captured 
by the simplified elastoplastic model used in this study. Also, crack propagation in crystalline quartz may differ from that of amorphous silica glass.

(b) For PMMA sliding on the tile surface, the theoretical model predicts that the removal of a single wear particle requires a large number (on the order of a few hundred) of asperity contacts. Due to the relatively short sliding distances employed in the experiments, it is unclear whether such a number of contacts actually occurred. As discussed in the main text, this suggests that the relationship between the tearing energy $\gamma$ and the crack tip displacement $\Delta x$ (i.e., the Paris curve) may require reinterpretation. In particular, crack propagation may not proceed in a continuous manner but may instead occur through stochastic, thermally assisted bond-breaking events. In this scenario, the crack tip may undergo finite, and potentially large, displacements in some asperity contacts, while in others no propagation occurs.

Paris curves are usually measured for relatively large, macroscopic samples (approximately $1 \ {\rm cm}$ in length), whereas wear typically involves the removal of micrometer-sized particles through crack propagation. Measuring the $\Delta x(\gamma)$ relationship for micrometer-sized cracks is an important objective for improving our understanding of wear processes.

\vskip 0.3cm
{\bf Declarations}

\vskip 0.2cm
{\bf Ethics approval and consent to participate: }
Not applicable.

\vskip 0.2cm
{\bf Consent for publication: }
Not applicable.

\vskip 0.2cm
{\bf Funding:}
Not applicable.

\vskip 0.2cm
{\bf Authors' contributions: }
All authors contributed equally to the work. All authors read and approved the final manuscript.

\vskip 0.2cm
{\bf Acknowledgements: }
We thank M. Ciavarella for useful discussions.

\vskip 0.2cm
{\bf Data availability: }
The data that support the findings of this study are available within the article. Additional data are available from the corresponding author upon reasonable request.


\begin{figure}
\includegraphics[width=0.47\textwidth,angle=0.0]{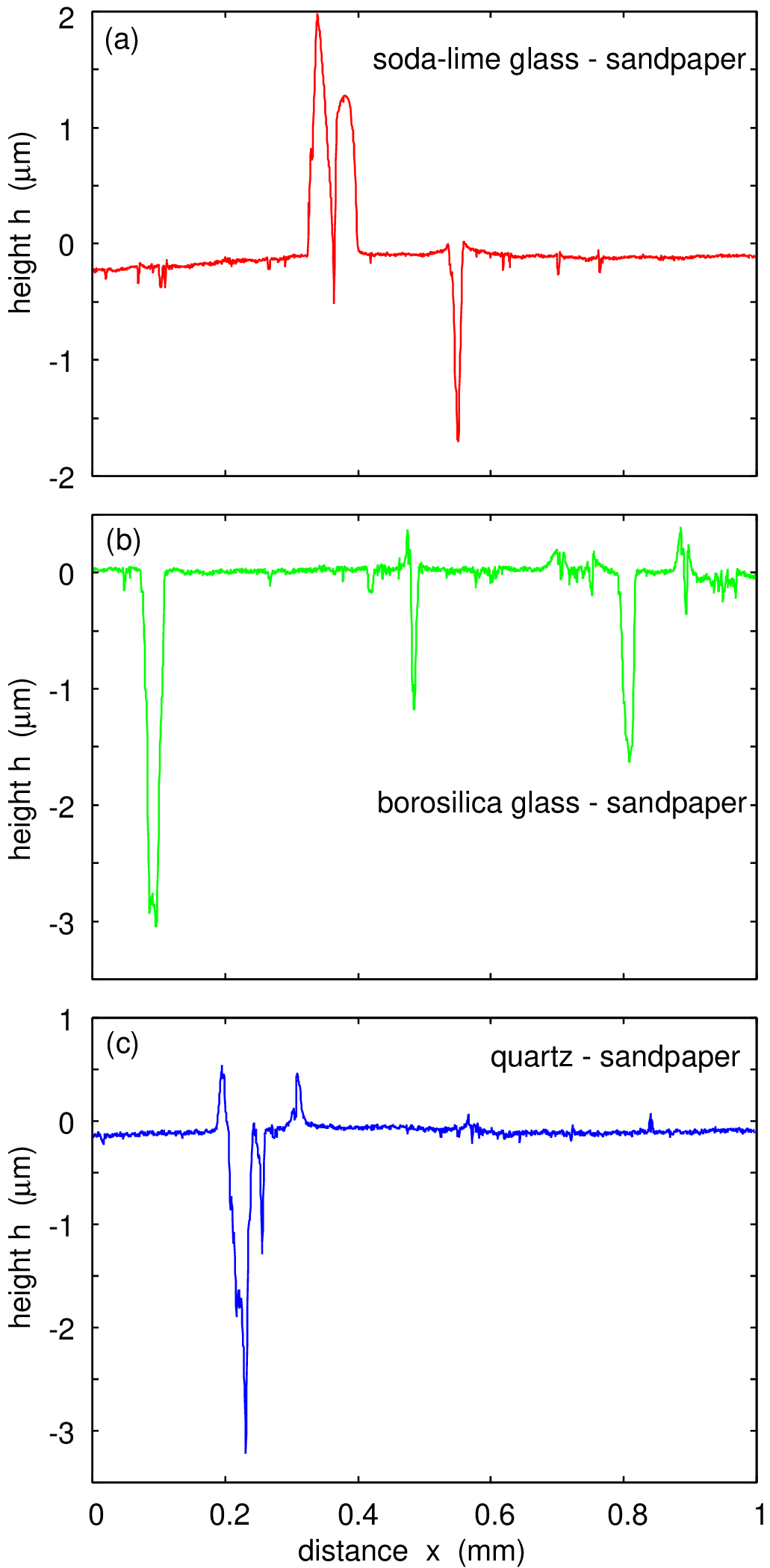}
\caption{\label{1x.2h.window.boro.quartz.eps}
The height profile orthogonal to the wear or ploughing tracks after a soda-lime glass block (a),
boro-silica glass block (b), and a quartz block (c) was slid 
a short distance on a sandpaper P100 surface. 
}
\end{figure}

\vskip 0.3cm
{\bf Appendix A} 

We have studied the ploughing tracks on  
a soda-lime glass block (a),  
a borosilicate glass block (b), and a quartz block (c) after they were slid  
a short distance on a sandpaper P100 surface (see Fig. \ref{1x.2h.window.boro.quartz.eps}).  
Even for these brittle materials, some of the material  
removed from below the undeformed surface plane is displaced by plastic flow to regions above the  
undeformed surface plane.

\end{document}